\documentclass[%
 reprint,
 amsmath,amssymb,
 aip,
 jcp,
]{revtex4-1}

\usepackage{amsmath}
\usepackage{graphicx}
\usepackage{dcolumn}
\usepackage{bm}
\usepackage{epsfig,color,xspace,multirow,xr,bbold}
\usepackage[all]{xy}
\usepackage{setspace}
\usepackage{url}

\usepackage{hyperref} 

\hypersetup{
   colorlinks,
   menucolor=blue,
   linkcolor=black,
   citecolor=blue,
   urlcolor=blue
}

\begin{document}

\title{Non-covalent interactions across organic and biological subsets
  of chemical space: Physics-based potentials parametrized from
  machine learning}

\author{Tristan Bereau}
\email{bereau@mpip-mainz.mpg.de}
\affiliation{Max Planck Institute for Polymer Research, 
Ackermannweg 10, 55128 Mainz, Germany}
\author{Robert A.~DiStasio Jr.}
\affiliation{Department of Chemistry and Chemical Biology, Cornell University,
  Ithaca, New York 14853, United States}
\author{Alexandre Tkatchenko}
\affiliation{Physics and Materials Science Research Unit, University of
  Luxembourg, L-1511 Luxembourg, Luxembourg}
\author{O.~Anatole von Lilienfeld}
\affiliation{Institute of Physical Chemistry and National Center for
  Computational Design and Discovery of Novel Materials (MARVEL), Department
  of Chemistry, University of Basel, Klingelbergstrasse 80, CH-4056 Basel,
  Switzerland}

\date{\today}

\begin{abstract}
  Classical intermolecular potentials typically require an extensive
  parametrization procedure for any new compound considered.  To do
  away with prior parametrization, we propose a combination of
  physics-based potentials with machine learning (ML), coined IPML,
  which is transferable across small neutral organic and
  biologically-relevant molecules.  ML models provide on-the-fly
  predictions for \emph{environment-dependent local} atomic
  properties: electrostatic multipole coefficients (significant error
  reduction compared to previously reported), the population and decay
  rate of valence atomic densities, and polarizabilities across
  conformations and chemical compositions of H, C, N, and O atoms.
  These parameters enable accurate calculations of intermolecular
  contributions---electrostatics, charge penetration, repulsion,
  induction/polarization, and many-body dispersion.  Unlike other
  potentials, this model is transferable in its ability to handle new
  molecules and conformations without explicit prior parametrization:
  All local atomic properties are predicted from ML, leaving only
  eight global parameters---optimized once and for all across
  compounds.  We validate IPML on various gas-phase dimers at and away
  from equilibrium separation, where we obtain mean absolute errors
  between 0.4 and 0.7~kcal/mol for several chemically and
  conformationally diverse datasets representative of non-covalent
  interactions in biologically-relevant molecules.  We further focus
  on hydrogen-bonded complexes---essential but challenging due to
  their directional nature---where datasets of DNA base pairs and
  amino acids yield an extremely encouraging 1.4~kcal/mol error.
  Finally, and as a first look, we consider IPML for denser systems:
  water clusters, supramolecular host-guest complexes, and the benzene
  crystal.
\end{abstract} 

\maketitle

\section{Introduction}

Our understanding of the physical laws that govern molecular
interactions have led to an ever-improving description of the
high-dimensional potential energy surface of condensed molecular
systems.  A variety of computational methods provide various
approximations thereof: while high-level methods (e.g., coupled
cluster) are restricted to a small number of atoms, other
electronic-structure methods (e.g., density functional theory---DFT)
can reach larger system sizes of up to $10^2-10^3$ atoms.  Beyond this
limit, classical potentials and force fields provide a much faster
estimate of the interactions, enabling the calculation of
thermodynamic and even kinetic properties for complex materials.

Many classical potentials and force fields are often termed
\emph{physics-based} because they encode assumptions about the
governing physics of the interactions via their functional forms.
Despite their widespread interest by the community, classical
potentials are currently limited to a narrow set of molecules and
materials, due to tedious and non-systematic parametrization
strategies.  Additive (i.e., non-polarizable) atomistic force fields
are typically parametrized from a combination of ab initio
calculations and experimental measurements, e.g., pure-liquid density,
heat of vaporization, or NMR chemical shifts.  Ensuring the accurate
reproduction of various molecular properties, from conformational
changes to thermodynamic properties (e.g., free energy of hydration),
but also consistency across all other molecules parametrized remains
challenging, time consuming, and difficult to automate.

Recently, a number of studies have brought forward the idea of more
automated parametrizations.  For instance, QMDFF is based on reference
DFT calculations to parametrize a set of classical
potentials.\cite{grimme2014general} We also point out the automatic
generation of intermolecular energies\cite{metz2016automatic}
extracted from reference symmetry-adapted perturbation
theory\cite{jeziorski1994perturbation} (SAPT) calculations.
Interestingly, recent efforts have aimed at parametrizing potentials
and force fields from atom-in-molecule (AIM) properties.  Van Vleet
\emph{et al.}\cite{van2016beyond} and Vandenbrande \emph{et
  al.}\cite{Vandenbrande2017} showed that a systematic use of AIMs can
significantly reduce the number of global parameters to scale the
individual energetic contributions.  Overall, they propose AIMs as a
means to more systematically parametrize models.  Similar conclusions
were reached for the additive OPLS force
field,\cite{cole2016biomolecular} for which the missing polarization
effects make a systematic scheme all the more challenging.  These
methodologies still require a number of a priori reference
electronic-structure calculations to optimize various parameters of
any new molecule encountered.

In the context of developing classical potentials for \emph{in silico}
screening across large numbers of compounds, the necessary
computational investment for the parametrization procedures of each
new molecule can become daunting.  A radically different strategy
consists in \emph{predicting} the potential energy surface of a system
from machine learning (ML).\cite{bartok2010gaussian, li2015molecular,
  behler2016perspective} ML encompasses a number of statistical models
that improve their accuracy with data.  Recent studies have reported
unprecedented accuracies in reproducing reference energies from
electronic-structure calculations, effectively offering a novel
framework for accurate intramolecular interactions freed from
molecular-mechanics-type approximations (e.g., harmonic
potential).\cite{chmiela2017machine, botu2016machine,
  schutt2017quantum} While they do away with free parameters that need
optimization (i.e., unlike force fields), they typically suffer from
limited transferability: an ML model is inherently limited to
interpolating across the training samples.  A model trained on water
clusters can be remarkably accurate toward describing liquid-state
properties (e.g., pair-correlation functions), but remains specific to
interactions solely involving water.\cite{natarajan2015representing}
Transferability of an ML model that would predict interactions across
chemical compound space (i.e., the diversity of chemical compounds)
stands nowadays as computationally intractable.  Part of the reason is
the necessity to interpolate across all physical phenomena for any
geometry, as these models are driven by experience, rather than
physical principles.  Symmetries and conservation laws will require
large amounts of data to be appropriately satisfied, if they are not
correctly encoded a priori.

In this work, we propose a balance between the aforementioned
physics-based models and an ML approach, coined IPML.  To best take
advantage of both approaches, we choose to rely on a physics-based
model, where most parameters are predicted from ML.  This approach
holds two main advantages: ($i$) Leverage our understanding of the
physical interactions at hand, together with the associated symmetries
and functional forms, and ($ii$) Alleviate the reference calculations
necessary to optimize the parameters of each new molecule.

The aforementioned AIM-based classical potentials, in this respect,
offer an interesting strategy: they largely rely on perturbation
theory to treat the long-range interactions (i.e., electrostatics,
polarization, and dispersion), while overlap models of
spherically-symmetric atomic densities describe the short-range
interactions.  Both theoretical frameworks estimate interaction
energies from \emph{monomer} properties---thereby significantly
reducing the ML challenge from learning interactions between any
combination of molecules to the much simpler prediction of (isolated)
atomic properties.  Incidentally, learning atomic and molecular
properties have recently been the subject of extended research,
providing insight into the appropriate representations and ML
models.\cite{rupp2012fast, hansen2013assessment,
  ramakrishnan2017machine, schutt2017quantum} Parametrizing
small-molecule force fields based on ML has already shown advantageous
at a more coarse-grained resolution.\cite{bereau2015automated} At the
atomistic level, Bereau \emph{et al.} had shown early developments of
learning AIM properties, namely distributed multipole coefficients to
describe the electrostatic potential of a
molecule.\cite{bereau2015transferable} The study was aiming at an
accurate prediction of multipole coefficients across the chemical
space of small organic molecules.  These coefficients provide the
necessary ingredients to compute the electrostatic interaction between
molecules via a multipole expansion.\cite{stone2013theory} Here, we
extend this idea by further developing physics-based models
parametrized from ML to all major interaction contributions:
electrostatics, polarization, repulsion, and dispersion.  We base our
method on a few ML models of AIM properties: distributed multipoles,
atomic polarizabilities from Hirshfeld ratios, and the population and
decay rate of valence atomic densities.  The combination of
physics-based potentials and ML reduces the number of global
parameters to only 7 in the present model.  We optimize our global
parameters once and for all, such that a new compound requires no
single parameter to be optimized (because the ML needs no refitting),
unlike most other aforementioned AIM- and physics-based
models.\cite{grimme2014general, metz2016automatic, van2016beyond}
Vandenbrande \emph{et al.}~did present results using frozen global
parameters, but their model still requires quantum-chemistry
calculations on every new compound to fit certain parameters (e.g.,
point charges).\cite{Vandenbrande2017} After parametrization on parts
of the S22x5 small-molecule dimer dataset,\cite{jurevcka2006benchmark}
we validate IPML on more challenging dimer databases of small
molecules, DNA base pairs, and amino-acid pairs.  We later discuss
examples beyond small-molecule dimers toward the condensed phase:
water clusters, host-guest complexes, and the benzene crystal.

\section{IPML: Physics-based potentials parametrized from machine learning}

\subsection{Learning of environment-dependent local atomic properties}

The set of intermolecular potentials is based on ML of local (i.e.,
atom in molecule) properties targeted at predicting electrostatic
multipole coefficients, the decay rate of atomic densities, and atomic
polarizabilities, which we present in the following.

\subsubsection{Electrostatic multipole coefficients}

The prediction of atomic multipole coefficients up to quadrupoles was
originally presented in Bereau \emph{et
  al.}\cite{bereau2015transferable} DFT calculations at the M06-2X
level\cite{zhao2008m06} followed by a GDMA
analysis\cite{stone2013theory} (i.e., wave-function partitioning
scheme) provided reference multipoles for several thousands of small
organic molecules.  ML of the multipoles was achieved using
kernel-ridge regression.  The geometry of the molecule was encoded in
the Coulomb matrix,\cite{rupp2012fast} ${\bf C}$, such that for two
atoms $i$ and $j$
\begin{equation}
  \label{eq:cmat}
  C_{ij} =
  \begin{cases}
    Z_i^{2.4}/2 & i =j, \\
    Z_iZ_j/r_{ij} & i \ne j.
  \end{cases}
\end{equation}
Though the Coulomb matrix accounts for translational and rotational
symmetry, it does not provide sufficient information to unambiguously
encode non-scalar, orientation-dependent quantities, such as dipolar
(i.e., vector) and quadrupolar (i.e., second-rank tensor) terms. A
consistent encoding of these terms had been achieved by rotating them
along a local axis system, provided by the molecular moments of
inertia.  To improve learning, the model aimed at predicting the
difference between the reference GDMA multipoles and a simple
physical, parameter-free baseline that helped identify symmetries in
vector and tensor components (hereafter mentioned as delta
learning). The large memory required to optimize kernel-ridge
regression models led us to construct one ML model per chemical
element.

In this work, we both simplify the protocol and significantly improve
the model's accuracy.  Reference multipoles are now extracted from DFT
calculations at the PBE0 level.  Rather than using GDMA multipoles, we
now rely on the minimal basis iterative stockholder (MBIS)
partitioning scheme.  While Misquitta \emph{et al.} recently
recommended the use of the ISA
multipoles,\cite{misquitta2014distributed} we use MBIS multipoles for
their consistency with the abovementioned atomic-density parameters
and the small magnitude of the higher multipoles, easing the learning
procedure.  We have also found MBIS multipoles to yield reasonable
electrostatic energies at long ranges (data not shown).  MBIS
multipoles were computed using {\sc Horton}.\cite{horton} Instead of
relying on the molecular moments of inertia as a local axis system, we
project each non-scalar multipole coefficient into a basis set $\{{\bf
  e}_{ij}, {\bf e}_{ik}, {\bf e}_{il}\}$ formed by three non-collinear
vectors ${\bf e}$ from the atom of interest $i$ to its three closest
neighbors: $j$, $k$, and $l$ (e.g., ${\bf e}_{ij} = ({\bf r}_j - {\bf
  r}_i)/|{\bf r}_j - {\bf r}_i|$, where ${\bf r}_i$ denotes the
Cartesian coordinates of atom $i$).  The vectors ${\bf e}_{ik}$ and
${\bf e}_{il}$ are further adjusted to form a right-handed orthonormal
basis set.

Further, the representation used for the ML model of electrostatic
multipoles is now the atomic Spectrum of London and
Axilrod-Teller-Muto potentials (aSLATM).\cite{huang2016communication,
  huang2017chemical} aSLATM represents an atomic sample and its
environment through a distribution of ($i$) chemical elements, ($ii$)
pairwise distances scaled according to London dispersion, and ($iii$)
triplet configurations scaled by the three-body Axilrod-Teller-Muto
potential.  We point out that aSLATM is atom-index invariant, and as
such does not suffer from discontinuities other representations may
have.  We used the QML implementation.\cite{qml} Point charges are
systematically corrected so as to yield an exactly neutral molecule.

\subsubsection{Atomic-density overlap}

Exchange-repulsion, as well as other short-ranged interactions, are
proportional to the overlap of the electron densities\cite{kim1981dependence,
  van2016beyond}
\begin{equation}
  S_{ij} = \int {\rm d}^3{\bf r}\  n_i({\bf r}) n_j({\bf r}).
\end{equation}
Van Vleet \emph{et al.}\cite{van2016beyond} presented a series of
short-ranged intermolecular potentials based on a Slater-type model of
overlapping valence atomic densities.  They approximated the atomic
density using the iterated stockholder atom (ISA)
approach\cite{lillestolen08redefining, misquitta2014distributed} The
atomic density of atom $i$, $n_i({\bf r})$, is approximated by a
single exponential function centered around the nucleus
\begin{equation}
  \label{eq:slater}
  n_i (r) \propto \exp(-\sigma_ir),
\end{equation}
where $\sigma_i$ characterizes the rate of decay of the valence atomic
density.  The short-ranged interactions proposed by Van Vleet \emph{et
  al.} rely on combinations of the decay rates of atomic densities,
i.e., $\sigma_{ij} = \sqrt{\sigma_i\sigma_j}$, for the atom pair $i$
and $j$.  While the decay rates were obtained from reference DFT
calculations, atom-type-dependent prefactors were fitted to
short-range interaction energies.  Vandenbrande \emph{et al.}~more
recently applied a similar methodology to explicitly include the
reference populations as normalization, $N_i = \int {\rm d}{\bf
  r}\ n_i ({\bf r}),$ i.e., the volume integrals of the valence atomic
densities.\cite{Vandenbrande2017} Their method allowed to reduce the
number of unknown prefactors per dimer: a single value for repulsion
and short-range polarization and no free parameter for penetration
effects (vide infra).

We constructed an ML model of $N$ and $\sigma$ using the same
representations and kernel as for Hirshfeld ratios (see above).
Reference coefficients $N$ and $\sigma$ were computed using {\sc
  Horton}\cite{horton, verstraelen2016minimal} for 1,102 molecules
using PBE0, amounting to 16,945 atom-in-molecule properties.  Instead
of the ISA approach, we followed Verstraelen \emph{et al.} and relied
on the MBIS partitioning method.\cite{verstraelen2016minimal}

\subsubsection{Atomic polarizabilities}

The Hirshfeld scheme provides a partitioning of the molecular charge
density into atomic contributions (i.e., an atom-in-molecule
description).\cite{hirshfeld1977bonded, tkatchenko2009accurate,
  bereau2014toward, buvcko2014extending} It consists of estimating the
change of atomic volume of atom $p$ due to the neighboring atoms, as
compared to the corresponding atom in free space
\begin{equation}
  \frac{V_p^{\rm eff}}{V_p^{\rm free}} = \frac{\int {\rm d}{\bf r}r^3
    w_p({\bf r})n({\bf r})}{\int {\rm d}{\bf r}r^3 n_p^{\rm free}({\bf
      r})},
\end{equation}
where $n_p^{\rm free}({\bf r})$ is the electron density of the free
atom, $n({\bf r})$ is the electron density of the molecule, and
$w_p({\bf r})$ weighs the contribution of the free atom $p$ against
all free atoms at ${\bf r}$
\begin{equation}
  w_p({\bf r}) = \frac{n_p^{\rm free}({\bf r})}{\sum_q n_q^{\rm
      free}({\bf r})},
\end{equation}
where the sum runs over all atoms in the
molecule.\cite{tkatchenko2009accurate} The static polarizability is
then estimated from the free-atom polarizability scaled by the
Hirshfeld ratio, $h$,\cite{gobre2016efficient}
\begin{equation}
  \alpha_p = \alpha_p^{\rm free} \left(\frac{V_p^{\rm eff}}{V_p^{\rm
      free}} \right)^{4/3} = \alpha_p^{\rm free} h^{4/3}.
\end{equation}

Reference Hirshfeld ratios were provided from DFT calculations of
1,000 molecules using the PBE0\cite{adamo1999toward} functional and
extracted using {\sc postg}.\cite{kannemann2010van, otero2013many} The
geometry of the molecule was encoded in the Coulomb matrix
(Eq.~\ref{eq:cmat}).  An ML model of the Hirshfeld ratios was built
using kernel-ridge regression and provided predictions for atomic
polarizabilities of atoms in molecules for the chemical elements H, C,
O, and N.  For all ML models presented here, datasets are split
between training and test subsets at an $80:20$ ratio, in order to
avoid overfitting.

\subsection{Intermolecular interactions from physics-based models}

In the following we present the different terms in our interaction
energy and how they rely on the abovementioned ML properties.

\subsubsection{Distributed multipole electrostatics}

The description of atom-distributed multipole electrostatics
implemented here follows the formalism of Stone.\cite{stone2013theory}
A Taylor series expansion of the electrostatic potential of atom $i$
gives rise to a series of multipole coefficients
\begin{align}
  \phi_i ({\bf r}) = \frac 1{4\pi\epsilon_0} &\left[ q_i \left(\frac 1r\right)
    - \mu_{i,\xi}\nabla_\xi \left(\frac 1r\right) \right. \notag\\
  & \left. + \frac 13 \Theta_{i,\xi\zeta} \nabla_\xi\nabla_\zeta
    \left(\frac 1r\right) - \ldots \right],
\end{align}
where $\xi$ and $\zeta$ indices run over coordinates and the
Einstein summation applies throughout.  We lump the multipole
coefficients in a vector $M_i = (q_i, \mu_{i,1}, \mu_{i,2}, \mu_{i,3},
\ldots)^t$ and derivatives of $1/r$ into the interaction matrix ${\bf
  T}^{ij} = (T^{ij},T_1^{ij},T_2^{ij},T_3^{ij},T_{11}^{ij},\ldots)^t$
for the interaction between atoms $i$ and $j$, where the number of
indices indicates the order of the derivative (e.g., $T_\xi^{ij} =
\nabla_\xi(1/r_{ij}))$.  In this way, the multipole electrostatic
interaction energy is given by
\begin{equation}
  \label{eq:mtp}
  E_{\rm elec} = \sum_{ij} M_iT^{ij} M_j.
\end{equation}
More details on the formalism and implementation of multipole
electrostatics can be found elsewhere.\cite{stone2013theory,
  ren2003polarizable, bereau2013leveraging} Multipole coefficients are
provided by the ML model for electrostatics originally presented in
Bereau \emph{et al.}\cite{bereau2015transferable} and improved herein
(see Methods above).

\subsubsection{Charge penetration}

The abovementioned multipole expansion explicitly assumes no
wavefunction overlap between molecules.  At short range, the
assumption is violated, leading to discrepancies in the electrostatic
energy, denoted \emph{penetration} effects.  The link between
penetration and charge-density overlap\cite{stone2013theory} has been
leveraged before by separating an atomic point charge into an
effective core and a damped valence electron
distribution.\cite{wang2010including, piquemal2003improved,
  wang2015general, narth2016scalable}.  An extension has later been
proposed by Vandenbrande \emph{et al.}  to efficiently estimate the
correction \emph{without} any free parameter.\cite{Vandenbrande2017}
This is achieved by including the atomic-density population $N_i$ of
atom $i$---the normalization term in Eqn.~\ref{eq:slater}.
Penetration is modeled by correcting the monopole-monopole
interactions in a pairwise fashion
\begin{align}
  \label{eq:sisa}
  E_{\rm pen} =& \sum_{ij} \frac{q_i^{\rm c}N_j}{r}g(\sigma_j, r)
  + \frac{N_iq_j^{\rm c}}{r}g(\sigma_i, r)\notag\\
  & -\frac{N_iN_j}r \left(f(\sigma_i,\sigma_j,r) + f(\sigma_j,\sigma_i,r)
  \right) \notag\\ 
  g(\sigma,r) =& \left(1+\frac r{2\sigma}\right) \exp\left(-\frac
    r{\sigma}\right),\notag\\ 
  f(\sigma_i,\sigma_j,r) =& \frac{\sigma_i^4}{(\sigma_i^2-\sigma_j^2)^2}
  \times \notag\\
  &\left(1 + \frac r{2\sigma_i} - \frac{2\sigma_j^2}{\sigma_i^2-\sigma_j^2}
  \right) \exp\left(-\frac r{\sigma_i}\right).
\end{align}
The present expression for $f(\sigma_i,\sigma_j,r)$ is problematic
when $\sigma_i \approx \sigma_j$ given the denominator, but
Vandenbrande \emph{et al.} derived corrections for such
cases.\cite{Vandenbrande2017} The parameter $q^c$ corresponds to a
core charge that is not subject to penetration effects, i.e., $q = q^c
- N$, where $q$ is determined from the multipole expansion.

We note the presence of three terms when considering electrostatics
together with penetration (Eqn.~\ref{eq:sisa}): the core-core
interaction (part of $E_{\rm elec}$, Eqn.~\ref{eq:mtp}), the damping
term between core and smeared density, and the last is the overlap
between two smeared density distributions.  In most existing
approaches, the damping functions aim at modeling the outer
Slater-type orbitals of atoms---e.g., note the presence of exponential
functions in Eqn.~\ref{eq:sisa}.  Unfortunately, penetration effects
due to the higher moments are not presently corrected.  Conceptually,
a separation between core and smeared contributions of higher
multipoles is unclear.  Rackers \emph{et al.}~proposed an interesting
framework that assumes a simplified functional form for the damping
term and factors out of the entire interaction matrix
$T^{ij}_\xi$.\cite{rackers2017optimized} We have not attempted to
express Eqn.~\ref{eq:sisa} for the interaction matrix $T^{ij}_\xi$
of all multipoles.

\subsubsection{Repulsion}

Following Vandenbrande \emph{et al.},\cite{Vandenbrande2017} we parametrize
the repulsive energy based on the overlap of valence atomic densities:
\begin{align}
  \label{eq:rep}
  E_{\rm rep} =& U^{\rm rep}_i U^{\rm rep}_j \sum_{ij} \frac{N_iN_j}{8\pi r} \left(
    h(\sigma_i, \sigma_j, r) + h(\sigma_j, \sigma_i, r)\right), \notag \\
  h(\sigma_i, \sigma_j, r) =& \left(
    \frac{4\sigma_i^2\sigma_j^2}{(\sigma_j^2-\sigma_i^2)^3} +
    \frac{\sigma_i}{(\sigma_j^2-\sigma_i^2)^2} \right) \exp\left( -\frac
    r{\sigma_i}\right),
\end{align}
where, $U^{\rm rep}_i$ is an overall prefactor that depends only on
the \emph{chemical element} of $i$.  The multiplicative mixing rule we
apply leads to $U^{\rm rep}_i$ having units of (energy)$^{1/2}$.  Here
again, corrections for $h(\sigma_i, \sigma_j, r)$ when $\sigma_i
\approx \sigma_j$ can be found elsewhere.\cite{Vandenbrande2017}

\subsubsection{Induction/polarization}

Polarization effects are introduced via a standard Thole-model description.
\cite{thole1981molecular} Induced dipoles, $\mu^{\rm ind}$, are
self-consistently converged against the electric field generated by both
multipoles and the induced dipoles themselves
\begin{equation}
  \mu^{\rm ind}_{i,\xi} = \alpha_i \left( \sum_{j} T^{ij}_\xi M_j +
    \sum_{j'} T^{ij'}_{\xi\zeta} \mu^{ind}_{j',\zeta} \right),
\end{equation}
where we follow the notation of Ren and
Ponder:\cite{ren2003polarizable} the first sum (indexed by $j$) only
runs over atoms \emph{outside} of the molecule containing $i$---a
purely intermolecular contribution---while the second sum (indexed by
$j'$) contains all atoms except for $i$.  We self-iteratively converge
the induced dipoles using an overrelaxation coefficient $\omega =
0.75$ as well as a smeared charge distribution, $n'$, following Thole's
prescription\cite{thole1981molecular} and the AMOEBA force
field\cite{ren2003polarizable}
\begin{equation}
  n' = \frac{3 a}{4 \pi} \exp\left( -au^3 \right),
\end{equation}
where $u = r_{ij}/(\alpha_i\alpha_j)^{1/6}$ and $a$ controls the
strength of damping of the charge distribution.  The smeared charge
distribution $n'$ leads to a modified interaction matrix, as described
by Ren and Ponder.\cite{ren2003polarizable} The electrostatic
contribution of the induced dipoles is then evaluated to yield the
polarization energy.  In this scheme, polarization thus relies on both
the predicted atomic polarizabilities and predicted multipole
coefficients.

\subsubsection{Many-body dispersion}

Many-body dispersion\cite{hermann2017first} (MBD) relies on the
formalism of Tkatchenko and coworkers.\cite{tkatchenko2012accurate} It
consists of a computationally efficient cast of the random-phase
approximation into a system of quantum harmonic
oscillators.\cite{donchev2006many}
In Appendix \ref{app:mbd} we briefly summarize the MBD implementation
and point the interested reader to Ref.~\citenum{bereau2014toward} for
additional details.

\subsubsection{Overall model}

To summarize, our intermolecular IPML model is made of five main
contributions: ($i$) electrostatics, ($ii$) charge penetration,
($iii$) repulsion, ($iv$) induction/polarization, and ($v$) many-body
dispersion.  Our use of ML to predict AIM properties yields only 8
global parameters to be optimized: ($i$) None; ($ii$) None; ($iii$)
$U^{\rm rep}_{\rm H}$, $U^{\rm rep}_{\rm C}$, $U^{\rm rep}_{\rm N}$,
$U^{\rm rep}_{\rm O}$; ($iv$) $a$; and ($v$) $\beta$, $\gamma$, $d$.
We will optimize these parameters simultaneously across different
compounds to explore their transferability.

We provide a Python-based implementation of this work at
\url{https://gitlab.mpcdf.mpg.de/trisb/ipml} for download.  The ML
models relied on kernel ridge regression, implemented here using {\sc
  numpy} routines.\cite{walt2011numpy} Different atomic properties
were trained on different datasets.  These datasets are also provided
in the repository.  While a single training set for all properties
would offer more consistency, different properties require very
different training sizes to reach an accuracy that is satisfactory.
Molecular configurations were generated from {\sc smiles} strings
using {\sc Open Babel}.\cite{o2011open} These approximate
configurations were purposefully \emph{not} further optimized to
obtain a more heterogeneous training set of configurations, thereby
improving the interpolation of the ML.

\section{Training and parametrization of IPML}

We show the accuracy of the prediction of the multipole coefficients,
the Hirshfeld ratios, the atomic-density decay rates, followed by the
assessment of experimental molecular polarizabilities.  We then
parametrize the different terms of the intermolecular potentials
against reference total energies on parts of the S22x5 dataset and
validate it against various other intermolecular datasets.

\subsection{Training of multipole coefficients}

\begin{figure*}[htbp]
  \begin{center}
    \includegraphics[width=0.8\linewidth]{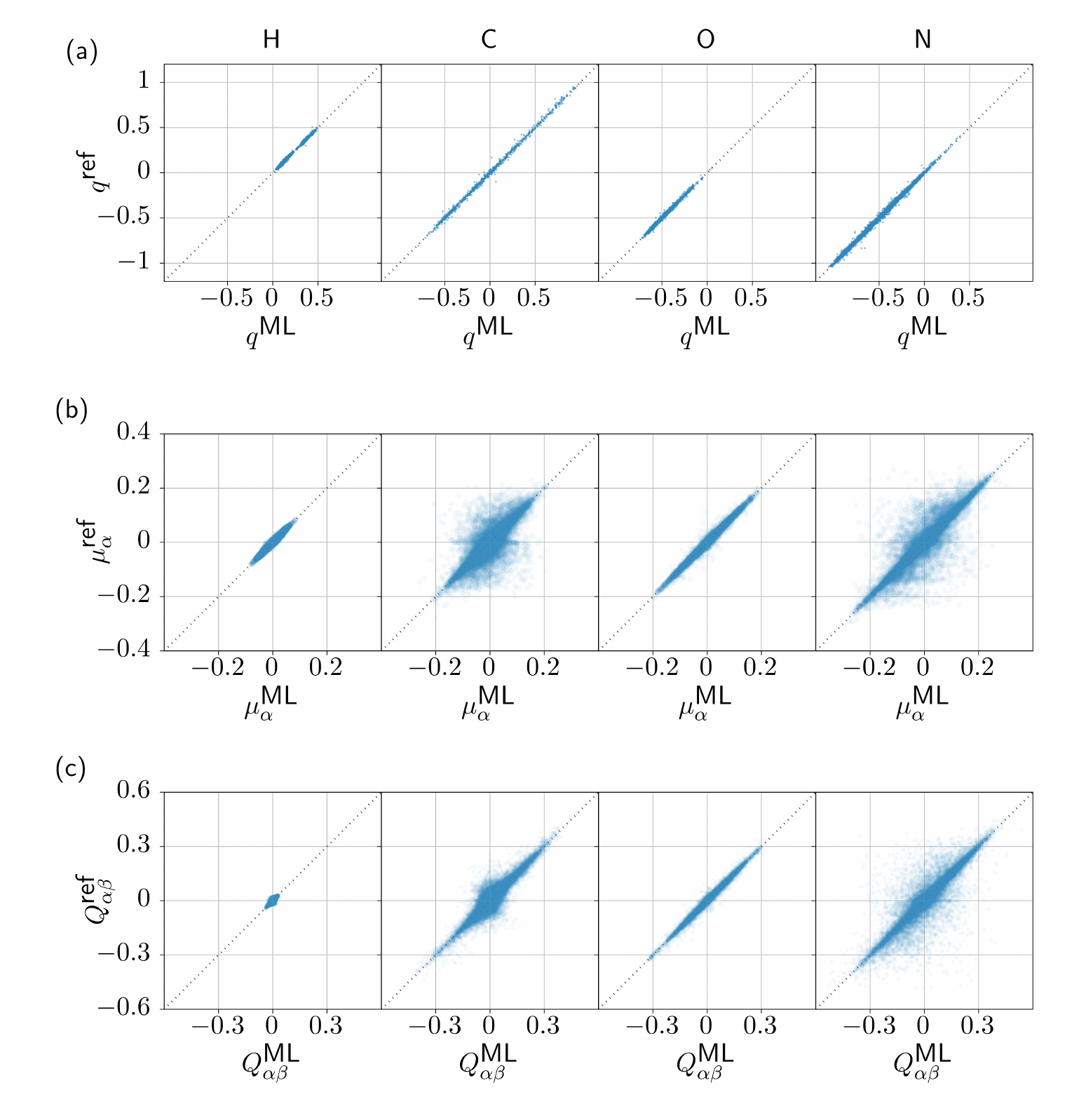}
    \caption{ML of the multipole coefficients of neutral molecules.
      Scatter correlation plots (out-of-sample predictions) for all
      components of (a) monopoles, (b) dipoles, and (c) quadrupoles of
      each chemical element, as predicted by the ML model with 80\%
      training fraction.  All quantities are expressed in units
      $e$\AA$^l$, where $l$ is the rank of the multipole.}
    \label{fig:mtp_corr}
  \end{center}
\end{figure*}

We performed ML of the multipole coefficients trained on up to 20,000
atoms in molecules---limited to neutral compounds.  While our
methodology allows us to learn all compounds together, we chose to
train an individual ML model for each chemical element.
Fig.~\ref{fig:mtp_corr} shows the correlation between reference and
predicted components for $\sim 10^4$ atoms in the test set.  Compared
to our previous report,\cite{bereau2015transferable} the accuracy of
the learning procedure is strongly improved for all ranks, i.e., MAEs
of $0.01~e$, $0.01~e\textup{\AA}$, and $0.02~e\textup{\AA}^2$ instead
of $0.04~e$, $0.06~e\textup{\AA}$, and $0.13~e\textup{\AA}^2$ for
monopoles, dipoles and quadrupoles, respectively.  The basis-set
projection used here yields significantly more accurate predictions
compared to the previously reported local-axis system augmented by a
delta-learning procedure.\cite{bereau2015transferable} We also point
out the strong improvement due to aSLATM (see below).  Finally, we
draw the reader's attention to the much smaller MBIS multipoles, as
compared to GDMA, thereby helping reaching lower MAEs.

\begin{figure}[htbp]
  \begin{center}
    \includegraphics[width=0.8\linewidth]{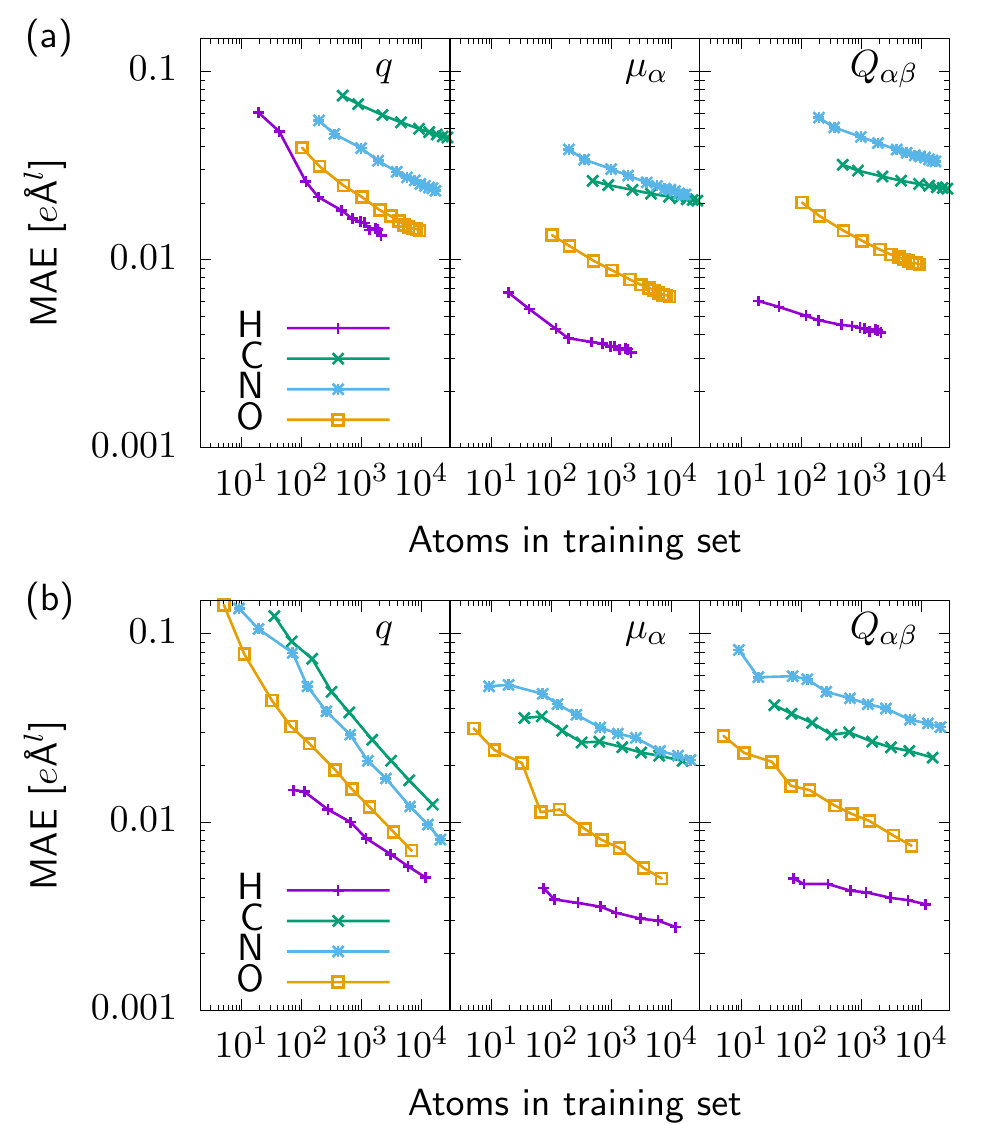}
    \caption{ML of the multipole coefficients of neutral
      molecules.  Comparison of representations: (a) Coulomb matrix
      and (b) aSLATM.  Saturation curves of the mean-absolute error
      (MAE) for monopoles, dipoles, and quadrupoles of each chemical
      element.  }
    \label{fig:mtp_sat}
  \end{center}
\end{figure}

Fig.~\ref{fig:mtp_sat} display learning curves for the different
multipole moments of each chemical element.  It compares the two
representations considered in this work: (a) Coulomb matrix and (b)
aSLATM.  The latter performs significantly better for point charges.
Though we reach excellent accuracy for the monopoles, some of the
higher multipoles remain more difficult, namely C and N.  On the other
hand, H and O both display excellent accuracy.  The main difference
between these two types of elements lies in their valency: H and O are
often found as terminal atoms, while N and C display much more complex
local environments.  This likely affects the performance of the
basis-set projection used in this work.  The similar learning
efficiency between the Coulomb matrix and aSLATM for dipoles and
quadrupoles further suggests the need for larger training sets (e.g.,
Faber \emph{et al.,} went up to 120,000
samples\cite{faber2017prediction}) or better local projections.  We
note the existence of ML methodologies that explicitly deal with
tensorial objects, though only applied to dipoles so
far.\cite{glielmo2017accurate, grisafi2017symmetry} In Appendix
\ref{app:mtp_cov}, we extend Glielmo \emph{et al.}'s covariant-kernel
description to quadrupoles using atom-centered Gaussian functions.
Tests on small training sets indicated results on par with
Fig.~\ref{fig:mtp_sat}.  We suspect that while covariant kernels offer
a more robust description of the rotational properties of tensorial
objects, the Coulomb matrix and aSLATM offer more effective
representations, offsetting overall the results.  Further, the
construction of covariant kernels is computationally involved: it
requires several outer products of rotation matrices to construct a $9
\times 9$ matrix (Eqns.~\ref{eq:appquadall} and \ref{eq:appquad}) for
a quadrupole alone.  This significant computational overhead led us to
use aSLATM with the basis-set projection for the rest of this work.
Covariant kernels for multipoles up to quadrupoles are nonetheless
implemented in our Python-based software.

\subsection{Training of valence atomic densities}

The accuracy of prediction of the populations and decay rates of
valence atomic densities, $N$ and $\sigma$, respectively, for a size
of the Coulomb matrix $n=6$ is shown on Fig.~\ref{fig:hirshfeld}a and
b.  The model was trained against 13,500 atoms in 800 molecules, and
tested against a separate set of 3,400 atoms in 200 molecules.  The
model shows high accuracy with MAEs of only 0.04~$e$ and 0.004~${\rm
  a.u.}^{-1}$, respectively.  Both models yield correlation
coefficients above 99.5\%.

\subsection{Training of Hirshfeld ratios}

\begin{figure}[htbp]
  \begin{center}
    \includegraphics[width=0.65\linewidth]{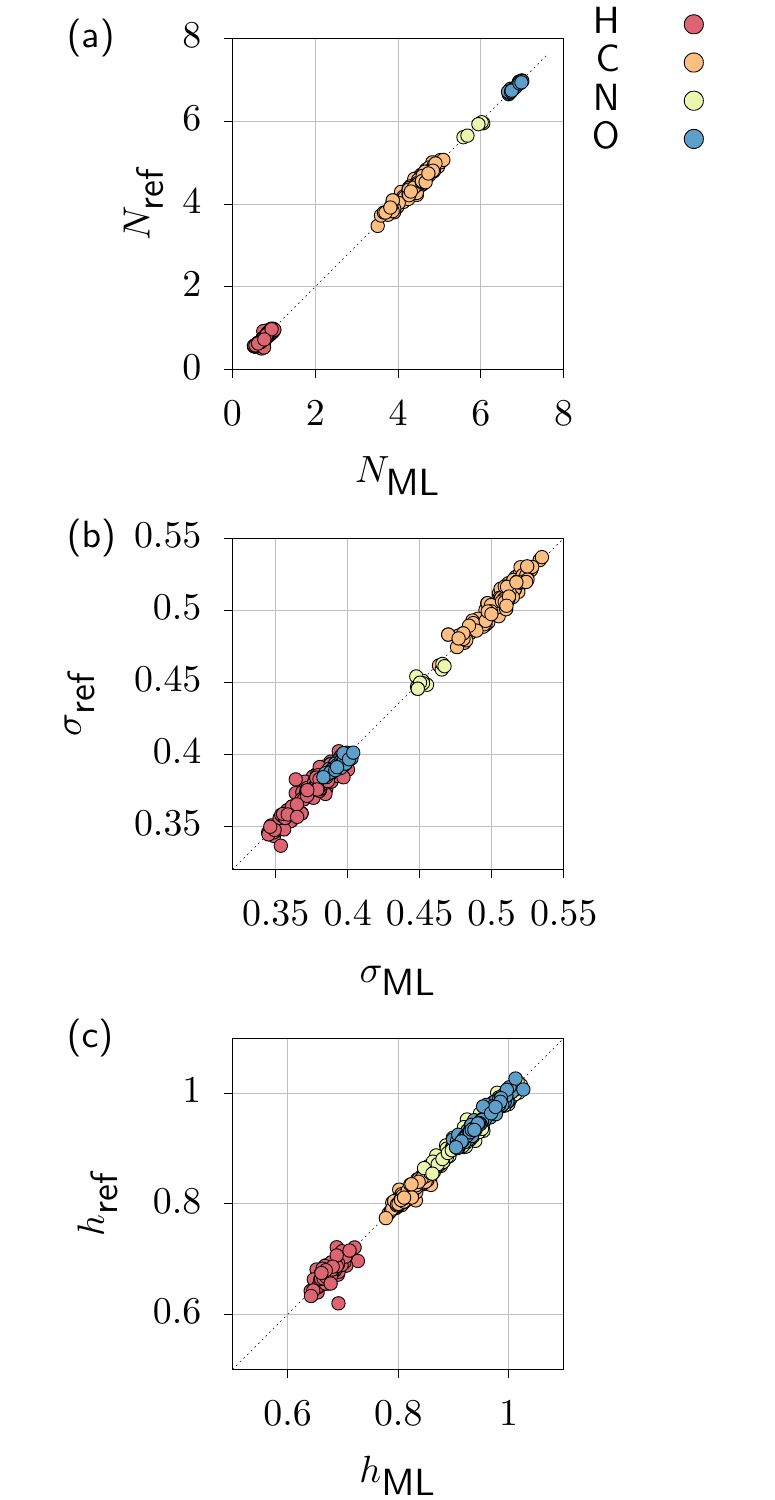}
    \caption{Correlation plots of out-of-sample predictions.  (a) ML
      of the populations (i.e., volume integral) of the valence atomic
      densities, $N$ (units in $e$).  (b) ML of the decay rate of the
      valence atomic densities, $\sigma$ (units in ${\rm a.u.}^{-1}$).
      (c) ML of the Hirshfeld ratios, $h$.  }
    \label{fig:hirshfeld}
  \end{center}
\end{figure}

Fig.~\ref{fig:hirshfeld}c shows a correlation plot of the predicted
and reference Hirshfeld ratios using the $n=12$ (i.e., size of the
Coulomb matrix) model trained against 12,300 atoms in 1,000 small
organic molecules.  We test the prediction accuracy on a different set
of 17,100 atoms.  We find high correlation (coefficient of
determination $R^2=99.5\%$) and a small MAE: 0.006.

\subsection{Molecular polarizabilities}

\begin{figure}[htbp]
  \begin{center}
    \includegraphics[width=\linewidth]{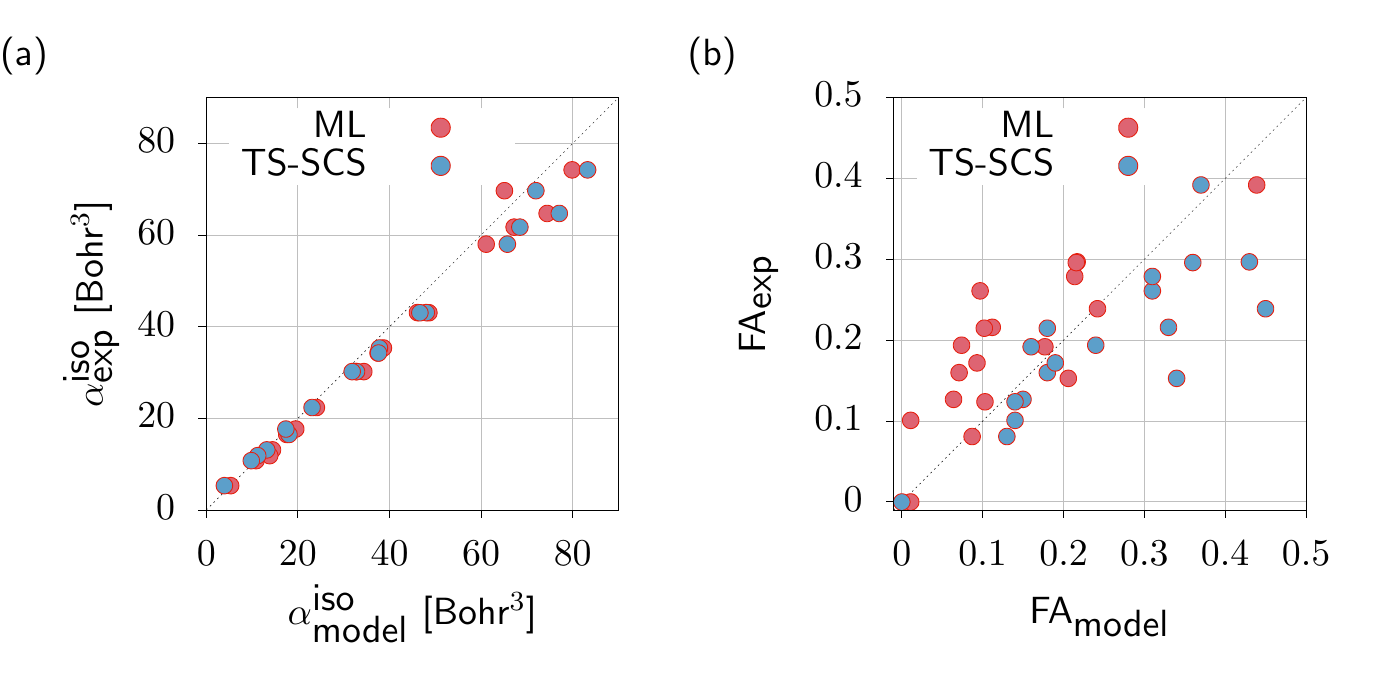}
    \caption{Correlation plot between (a) isotropic and (b) fractional
      anisotropies of molecular polarizabilities predicted from the
      current ML model (blue) and Tkatchenko-Scheffler
      polarizabilities after SCS
      procedure\cite{tkatchenko2009accurate, distasio2014many} (red)
      against experimental values for the set of 18 compounds proposed
      in Ref.~\citenum{thole1981molecular}.  }
    \label{fig:alphas}
  \end{center}
\end{figure}

Predictions of the Hirshfeld ratios were further assessed by
calculating (anisotropic) molecular polarizabilities.  Reference
experimental values of 18 small molecules were taken from
Thole,\cite{thole1981molecular} for both the isotropic molecular
polarizability as well as the fractional anisotropy, as defined
elsewhere.\cite{bereau2014toward} Fig.~\ref{fig:alphas} shows both the
isotropic (panel a) and fractional anisotropy (panel b), comparing the
present ML prediction with calculations using the Tkatchenko-Scheffler
method after solving the self-consistent screening (SCS)
equation.\cite{tkatchenko2009accurate, distasio2014many} We find
excellent agreement between the ML prediction and experiment for the
isotropic component: an MAE of $3.2$~Bohr$^3$ and a mean-absolute
relative error (MARE) of 8.6\%, both virtually identical to the
Tkatchenko-Scheffler calculations after SCS.\cite{distasio2014many}
The fractional anisotropy tends to be underestimated, though overall
the agreement with experiment is reasonable, as compared to previous
calculations that explicitly relied on DFT calculations for each
compound.

\subsection{Parametrization of the intermolecular energies}

To optimize the abovementioned free parameters, we aimed at
reproducing the intermolecular energies of a representative set of
molecular dimers.  The collection of global parameters optimized
during this work are reported in Tab.~\ref{tab:param}.  The
parameters, shown in Tab.~\ref{tab:param}, were optimized
simultaneously using basin hopping\cite{wales1997global,
  wales2003energy} to reproduce the total intermolecular energy from
reference calculations.  We also provide a rough estimate of the
sensitivity of these parameters through the standard deviation of all
models up to 20\% above the identified global minimum.  We introduce
chemical-element-specific prefactors for the repulsion interaction.
The repulsive interaction is thus scaled by the \emph{product} of
element specific prefactors for each atom pair.  The apparent lack of
dependence of the dispersion parameter $d$ led us to fix it to the
value $d=3.92$.\cite{bereau2014toward}

\begin{table*}[htbp]
  \begin{center}
    \begin{tabular*}
      {0.9\linewidth}{@{\extracolsep{\fill}} l|c||rr|rr}
      Interaction & Parameter & \multicolumn{2}{c|}{Model 1} & \multicolumn{2}{c}{Model 2} \\
      && value & sensitivity & value & sensitivity \\
      \hline 
      \hline
      Polarization   & $a$ & 0.0187 & 0.09 & 0.0193 & 0.03 \\
      Dispersion     & $\gamma$ & 0.9760 & 0.04 & 0.9772 & 0.04\\
      & $\beta$            & 2.5628 & 0.08 & 2.2789 & 0.04\\
      & $d$                & 3.92 & & 3.92 & \\
      Repulsion   & $U^{\rm rep}_{\rm H}$ & 27.3853 & 1 & 23.5936 & 1\\
      & $U^{\rm rep}_{\rm C}$ & 24.6054 & 0.5 & 24.0509 & 0.5\\
      & $U^{\rm rep}_{\rm N}$ & 22.4496 & 0.6 & 21.4312 & 0.3\\
      & $U^{\rm rep}_{\rm O}$ & 16.1705 & 0.8 & 16.0782 & 0.2\\
      \hline
    \end{tabular*}
    \caption{Optimized global parameters determined from two different
      training sets.  Model 1: Fitting to the S22x5 at distances 0.9x
      and 1.0x.  Model 2: Fitting to the S22x5 at distance 1.0x and
      S12L.  Parameters $U_{\rm X}$ correspond to the repulsion of
      chemical element X, expressed in (kcal/mol)$^{1/2}$.  ``Value''
      corresponds to the optimal parameter, while ``sensitivity''
      reflects the standard deviation of parameters around (up to 20\%
      above) the identified global minimum.  Sensitivity is not
      provided for $d$ (see main text).}
    \label{tab:param}
  \end{center}
\end{table*}

A better understanding of the variability of our global parameters led
us to consider two sets of reference datasets for fitting, coined
below model 1 and model 2.  While model 1 only considers
small-molecule dimers, model 2 also incorporates host-guest complexes.
For both models we rely on the S22x5 small-molecule
dataset\cite{jurevcka2006benchmark, grafova2010comparative} at the
equilibrium distance (i.e., 1.0x distance factor).  In addition, model
1 also considers configurations at the shorter distance factor 0.9x to
help improve the description of the curvature of the potential energy
landscape.  Model 2, on the other hand, adds to S22x5 at 1.0x a series
of host-guest complexes: the S12L database.\cite{ambrosetti2014hard}
All the results presented below will be derived from model 1, unless
otherwise indicated.  The comparison with model 2 aims at showing
($i$) the robustness of the fit from the relatively low variability of
global parameters (except possibly for $U_{\rm H}$) and ($ii$) an
outlook toward modeling condensed-phase systems.

\begin{figure}[htbp]
  \begin{center}
    \includegraphics[width=0.85\linewidth]{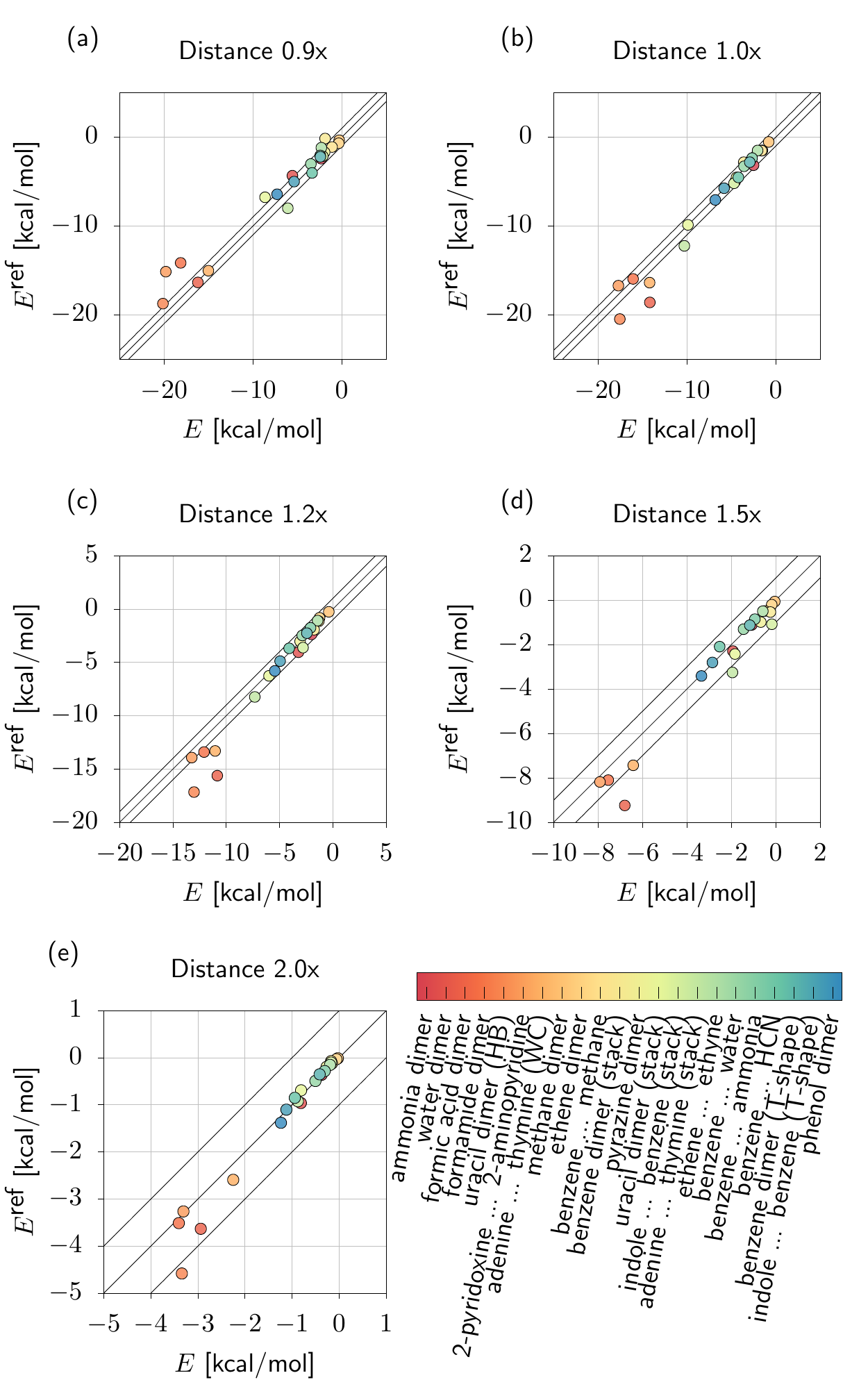}
    \caption{Correlation of intermolecular energies for S22x5. The
      different panels describe the interactions at specific distance
      factors (i.e., from 0.9x to 2.0x).  Color coding corresponds to
      the compound ID---hydrogen bonding compounds correspond to low
      values while van der Waals compounds correspond to the larger
      values.  The different diagonals bracket the $\pm 1$~kcal/mol
      area of accuracy.}
    \label{fig:s22x5}
  \end{center}
\end{figure}

While the overall MAE averaged over all distance factors is
0.7~kcal/mol, the error clearly drops with distance: 1.0, 0.8, 0.8,
0.5, and 0.2 for distance factors 0.9x, 1.0x, 1.2x, 1.5x, and 2.0x,
respectively.  This illustrates that the model yields robust
asymptotics, with significant improvement compared to a cruder model
that only included multipole electrostatics and many-body
dispersion.\cite{bereau2014toward} Outliers from the $\pm1~$kcal/mol
accuracy region are composed of strongly hydrogen-bonding complexes
(e.g., 2-pyridoxine with 2-aminopyridine), which depend significantly
on the quality of the electrostatic description.  The correlation
achieved here depends critically on the accuracy of the multipole
moments.  Indeed, the few global parameters included in our model
provide little room for error compensations.  For instance, we found
that a poorer ML model of the multipole moments yielded significant
artifacts on the partial charges of hydrogen cyanide, leading to an
artificially strong polarization of the hydrogen.

We also point out the small value of the polarization parameter, $a$
(Tab.~\ref{tab:param}), leading effectively to small polarization
energies.  Rather than an imbalance in the model, we suspect that
significant short-range polarization energy is absorbed in the
repulsion terms.  Indeed, several AIM- and physics-based force fields
use the same overlap model to describe repulsion and short-range
polarization.\cite{van2016beyond, Vandenbrande2017} Since we optimize
all terms directly against the total energy rather than decompose each
term, such cancellations may well occur.  We also expect that
including systems in which strong non-additive polarization effects
would play a role in outweighing effective pairwise polarization.  In
addition, we note that the pairwise scheme is optimized per chemical
element, while the Thole model is not.

\section{Performance of the IPML model}

\subsection{Non-equilibrium geometries (S66a8)}

\begin{figure}[htbp]
  \begin{center}
    \includegraphics[width=0.7\linewidth]{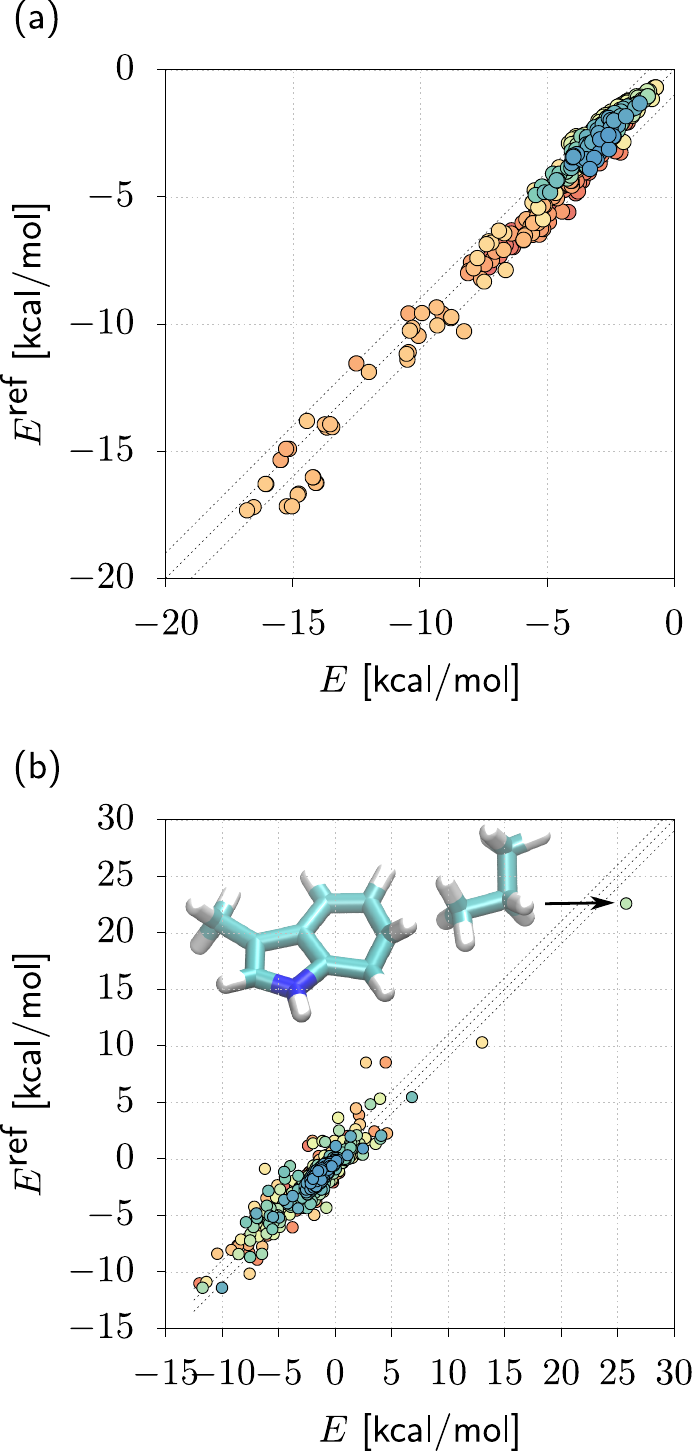}
    \caption{Correlation plots for the total intermolecular energy
      between reference and present calculations for (a) the S66a8
      dataset of dimers translated and rotated away from their
      equilibrium geometry and (b) the SSI dataset of amino acids
      (only dimers involving neutral compounds made of HCON atoms).
      Inset: strongly-repulsive tryptophan-glutamine dimer.}
    \label{fig:s66ax}
  \end{center}
\end{figure}

A recent extension of the S66 dataset of molecular dimers provides
angular-displaced non-equilibrium geometries, i.e., S66a8
($66\times8=528$ dimers).\cite{rezac2011extensions} The correlation
between our model and reference calculations at the CCSD(T)/CBS level
of theory are presented in Fig.~\ref{fig:s66ax}a.  Excellent agreement
is found for most samples, with an MAE of only $0.4$~kcal/mol across a
larger, representative set of molecular dimers, as compared to the S22
used for training.  Model 2 performs virtually on par with an MAE of
only $0.5$~kcal/mol.  

We compare our results with the MEDFF model whose overlap model is
used in the present work, but relies on point-charge electrostatics
and a pairwise dispersion model.\cite{Vandenbrande2017} They report
root-mean squared errors of 0.36~kcal/mol for the dispersion-dominated
complexes of the S66 dataset at equilibrium distances.  Given that
hydrogen-bonded complexes are typically more
challenging,\cite{grimme2014general, Vandenbrande2017} our model
likely compares favorably, keeping in mind that the dataset and error
measurement are different.  They also report a reduced 0.26~kcal/mol
error over the entire S66 dataset when each parameter is optimized
specifically for each complex.  Given our focus on model
transferability, we did not attempt a similar measurement.  For the
same dataset and error measurement, the QMDFF model reports a larger
1.1~kcal/mol error.\cite{Vandenbrande2017}

\subsection{Amino-acid side chains (SSI dataset)}

The SSI dataset contains pairs of amino-acid side chains extracted
from the protein databank.\cite{sherrill2017biofragment} We removed
dimers containing charged compounds and sulfur-containing side chains
(i.e., cysteine and methionine), for a total of 2,216 dimers.  We
computed intermolecular energies using the present method and compare
them with reference CCSD(T) at the complete basis set limit.  In
Fig.~\ref{fig:s66ax}b we compare the total energy with reference
energies.  We find again excellent agreement throughout the much
larger range.  We note the presence of a high-energy dimer at
$+23$~kcal/mol, corresponding to a tryptophan-glutamine dimer (inset
of Fig.~\ref{fig:s66ax}b).  The strong deformation of the tryptophan
ring illustrates the robustness of our model in accurately reproducing
intermolecular interactions for a variety of conformers.  Model 1
yields overall an MAE of 0.37~kcal/mol.  Interestingly, this accuracy
is on par with additive force fields, such as GAFF and CGenFF (0.35
and 0.23~kcal/mol, respectively), and better than certain
semi-empirical methods, e.g., AM1
(1.45~kcal/mol).\cite{sherrill2017biofragment} Model 2 yields
virtually the same MAE, 0.38~kcal/mol, but underpredicts the
high-energy dimer highlighted in Fig.~\ref{fig:s66ax}b: 3.6 instead of
22.6~kcal/mol.  It highlights how widening the training set of the
model to both small molecules and host-guest complexes decreases the
accuracy on the former.

\subsection{DNA-base and amino-acid pairs (JSCH-2005)}

\begin{figure*}[htbp]
  \begin{center}
    \includegraphics[width=0.9\linewidth]{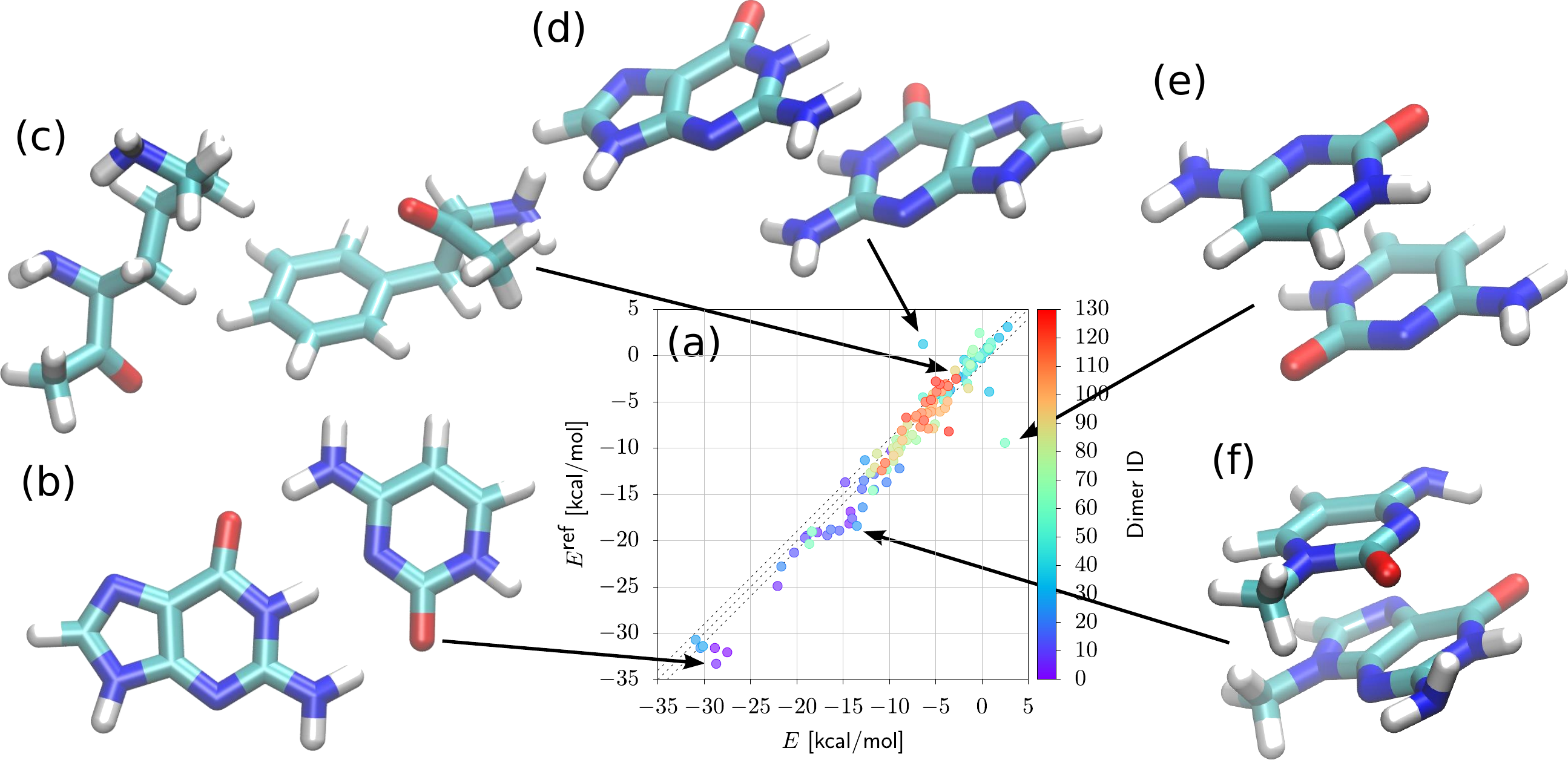}
    \caption{(a) Correlation plots for the total intermolecular energy between
      reference and present calculations for the JSCH-2005
      dataset\cite{jurevcka2006benchmark} of DNA-base and amino-acid pairs
      (dimers involving charged compounds are not shown). (b) GC in a
      Watson-Crick geometry; (c) Lysine and phenylalanine; (d) GG complex; (e)
      CC complex; (f) methylated GC complex.}
    \label{fig:jsch}
  \end{center}
\end{figure*}

The JSCH-2005 dataset offers a benchmark of representative DNA base
and amino-acid pairs.\cite{jurevcka2006benchmark} Again, we focus on
neutral molecules only, for a total of 127 dimers.  The correlation of
total interaction energies is shown in Fig.~\ref{fig:jsch}a.  We find
a somewhat larger MAE of 1.4~kcal/mol.  This result remains extremely
encouraging, given the emphasis of strong hydrogen-bonded complexes
present in this dataset.  While others have pointed out the challenges
associated with accurately modeling these
interactions,\cite{grimme2014general, Vandenbrande2017} we have not
found reference benchmarks on specific datasets such as this one for
similar physics-based models.  Given the prevalence of hydrogen bonds
in organic and biomolecular systems, we hope that this work will
motivate a more systematic validation on these interactions.

Representative examples are shown on Fig.~\ref{fig:jsch}.  While the
Watson-Crick complex of the guanine (G) and cytosine (C) dimer (panel
a) leads to one of the strongest binders, weak hydrogen bonds can
still lead to the dominant contribution, as seen in (f) for the
methylated GC complex.  We find two outliers, shown in (d) and (e),
where $\pi$-stacking interactions dominate the interaction energy.
The discrepancies likely arise from an inadequate prediction of some
quadrupole moments, especially involving nitrogen (see
Fig.~\ref{fig:mtp_corr}).  Note the structural similarity between (d),
(e), and (f): the weak hydrogen bonds in the latter case dominate the
interaction and resolve any apparent discrepancy with the reference
energy.  For this dataset, model 2 performs significantly worse, with
an MAE of $2.3$~kcal/mol, indicating that forcing transferability
across both small-molecule dimers and host-guest complexes strains the
accuracy of the model for challenging small molecules exhibiting
significant $\pi$-stacking and hydrogen-bonding behavior.  This
significant change in performance contrasts the very similar
parameters between the two models, highlighting a sensitive parameter
dependence.

\subsection{Water clusters}

\begin{figure}[htbp]
  \begin{center}
    \includegraphics[width=0.7\linewidth]{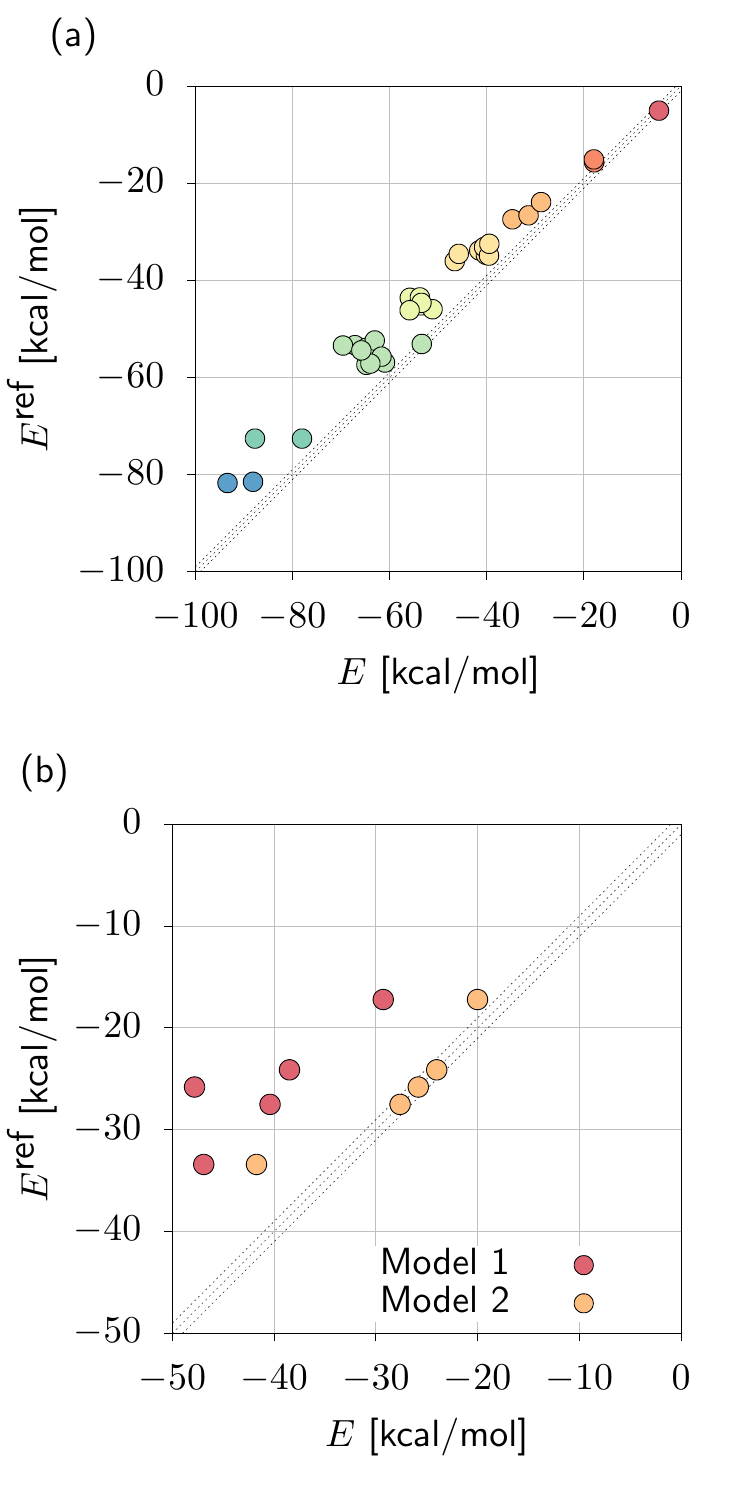}
    \caption{Correlation plots for the total intermolecular energy
      between reference and present calculations for (a) the
      water-clusters dataset and (b) the host-guest complexes in the
      S12L database.  The colors in (a) indicate the number of
      molecules involved in the cluster: from two (red) to 10 (blue)
      molecules.}
    \label{fig:watclust_s12l}
  \end{center}
\end{figure}

Beyond dimers, we test the ability of our potentials to reproduce
energies of larger clusters.  Fig.~\ref{fig:watclust_s12l}a shows the
correlation of the total energy between the present work and CCSD(T)
calculations at the complete basis set limit of water clusters
involving from 2 to 10 molecules.\cite{temelso2011benchmark} The
model's energies correlate highly with the reference but progressively
overstabilize.  This shift results from compounding errors that grow
with cluster size, amounting to an MAE of 8.1~kcal/mol.  Note that we
can correct the slope by including a single water cluster in the
above-mentioned parametrization (data not shown).  Model 2 performs
virtually on par with model 1.

IPML recovers the overall trend of energies for complexes of various
sizes, but there is still room for improvements.  This is notable
given that the many-body polarization term was optimized to zero in
both models (see Tab.~\ref{tab:param}).  It indicates that a pairwise
description captures the main effects even for the larger complexes
considered here.  Improving the results would require forcing the
parametrization to rely more significantly on many-body polarization.
Improving the modeling of other terms, such as repulsion, may also
help reduce incidental cancellations of errors.

\subsection{Supramolecular complexes (S12L)}

Moving toward more complex systems, we test the ability to reproduce
intermolecular energies of host-guest complexes.
Fig.~\ref{fig:watclust_s12l}b shows the correlation of the total
intermolecular energy against diffusion Monte
Carlo.\cite{ambrosetti2014hard} Although we find high correlation, the
MAE is substantial: 9.7~kcal/mol.  A comparison with model 2, which
significantly improves the agreement, demonstrates the benefit of
including larger complexes in the fit of the global parameters.
Still, one outlier remains: the glycine anhydride-macrocycle, with an
overstabilization of $8$~kcal/mol, despite being fitted into the
global parameters.  This compound (displayed in Fig.~8 of
Ref.~\citenum{bereau2014toward}) displays sites at which multiple
hydrogen bonds coincide.  It further suggests the role of inaccurate
multipoles, as well as an inadequate electrostatic penetration model
(i.e., missing higher-order multipoles beyond monopole correction),
and possibly many-body repulsion interactions.

\subsection{Benzene crystal}

\begin{figure}[htbp]
  \begin{center}
    \includegraphics[width=\linewidth]{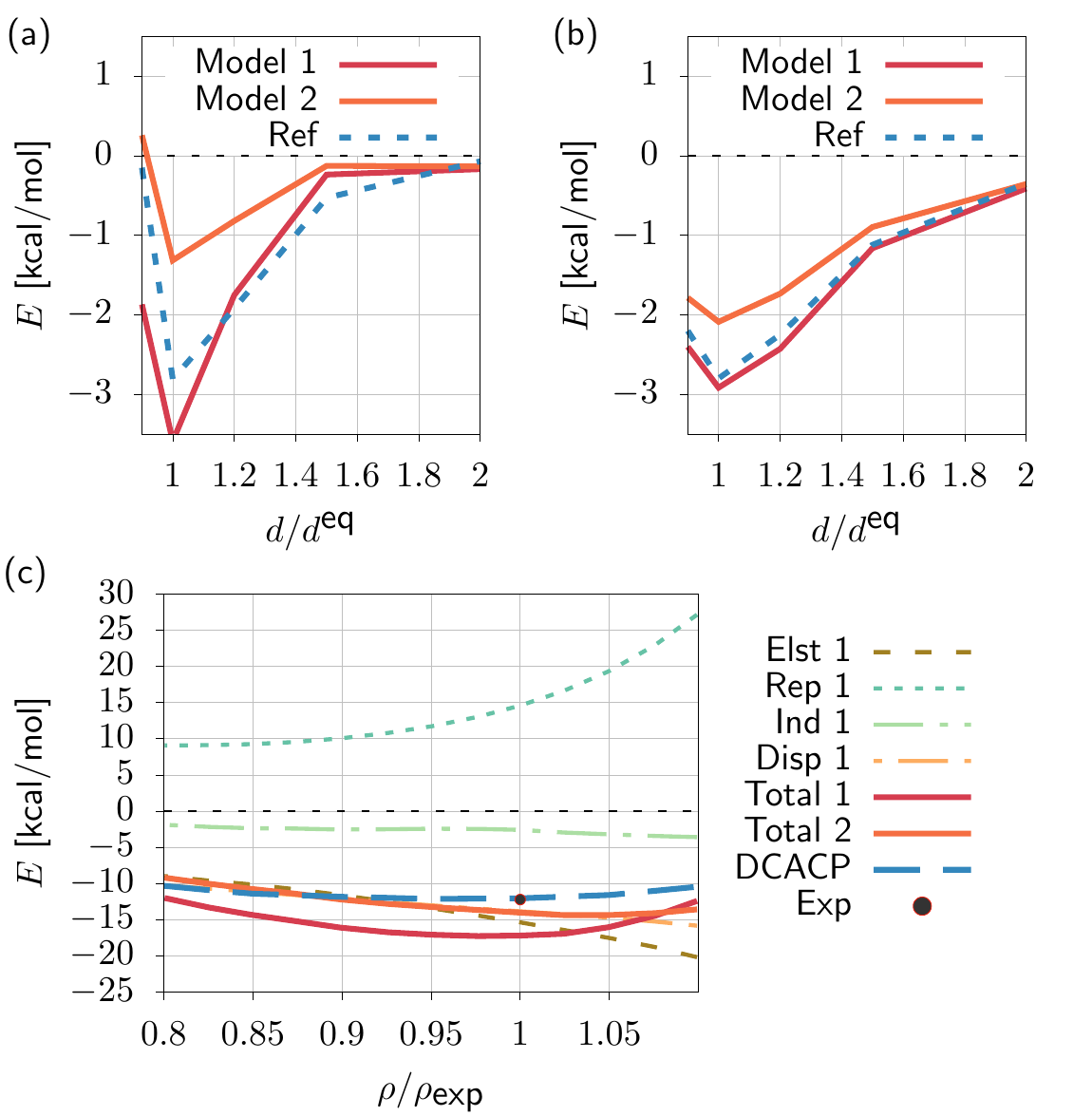}
    \caption{Comparison of the intermolecular energy as a function of
      dimer distance for the benzene dimer in the (a)
      parallel-displaced and (b) T-shaped conformations. (c) Cohesive
      binding energy of the benzene crystal as a function of the
      scaling factor of the unit cell.  }
    \label{fig:bzn_cryst}
  \end{center}
\end{figure}

As another example leading to condensed-phase properties, we evaluate
the model's ability to reproduce the cohesive energy of the benzene
crystal.  We scale the lattice unit cell around the equilibrium value,
as detailed in previous work.\cite{bereau2014toward} The various
contributions of the energy are shown in Fig.~\ref{fig:bzn_cryst}c.
For reference, we compare the cohesive energy with experimental
results\cite{schweizer2006quantum} and dispersion-corrected
atom-centered potentials (DCACP).\cite{tapavicza2007weakly}

As reported before,\cite{von2010two, bereau2014toward} we find the
benzene crystal to display significant dispersion interactions.
Though the overall curvature against density changes agrees reasonably
well with DCACP, we find that the method overstabilizes the molecular
crystal.  Model 1 yields a cohesive energy of $-17.2$~kcal/mol at
equilibrium, as compared to the experimental value of
$-12.2$~kcal/mol.\cite{schweizer2006quantum} For reference, we show
the potential energy landscapes of the benzene dimer in the stacked
(a) and T-shaped (b) conformations.  Excellent agreement is found in
the latter case, while the former shows an overstabilization.

Interestingly, while model 2 seems to understabilize these two dimer
configurations, it better reproduces the cohesive energy of the
crystal, with a value at equilibrium density of $-14.3$~kcal/mol, only
$2$~kcal/mol away from the experimental value.  We conclude that the
inclusion of host-guest complexes in the optimization of the global
parameters helps describe systems toward or in the condensed phase.
Still, the compounding errors present in the model limit a systematic
extension to molecular crystals.  We again point at the necessity for
extremely accurate multipole moments, where any discrepancy can have
significant effects in the condensed phase.  Further improving the
prediction of the multipole moments will strongly contribute to an
improved accuracy of the present energy model.




\section{Conclusions and future outlook}

We have presented a set of classical potentials to describe the
intermolecular interactions of small molecules, coined IPML.  Notably,
we present a methodology that readily provides parameters for a large
range of small molecules by relying on atom-in-molecule properties
predicted from machine learning (ML).  Predictions for distributed
multipoles, Hirshfeld ratios, valence atomic density decay rate and
population provide the necessary parameters for electrostatics,
polarization, repulsion, and many-body dispersion.  Remarkably, our
methodology provides a first attempt at transferable intermolecular
potentials with few global parameters optimized across a subset of
chemical space containing H, C, N, and O atoms only.  In contrast to
other studies, we do not reoptimize the global parameters for every
new compound.  We rationalize this by the use of more sophisticated
physical models, e.g., many-body rather than pairwise dispersion,
multipole rather than point-charge electrostatics, and non-additive
rather than pairwise additive polarization.

As compared to purely data-driven methodologies, IPML starts from
physics-based interactions and only relies on ML to predict parameters
thereof.  Perturbation theory and the short-range overlap method offer
an appealing framework to describe interactions based on monomer
properties---effectively simplifying greatly the training of ML models
of parameters.  Conceptually, blending physical constraints in a
data-driven framework would ideally translate into setting the
functional form of the interaction as a prior of the ML model.  As an
example, reproducing kernel Hilbert space can fit a potential energy
surface by imposing the asymptotics at long range.\cite{ho1996general,
  unke2017toolkit}

Extensions of the present work to a force field would amount to
computing derivatives.  Analytical derivatives of the potentials with
respect to atomic coordinates are either straightforward (e.g.,
pairwise repulsion and charge penetration) or already available (e.g.,
many-body dispersion\cite{blood2016analytical} or electrostatics and
induction\cite{ponder2010current}).  Our ML models being
conformationally dependent, computation of the forces would also
entail a derivative with respect to the atom-in-molecule properties.
While not implemented here, this information can readily be extracted
from derivatives of the kernel used in the
ML.\cite{rasmussen2006gaussian} How to optimize such a
conformationally-dependent force field to best balance the extra
accuracy with the additional computational overhead remains an open
problem.

Even though we did not aim at a performance optimization, the present
implementation can help us gain insight into the computational cost of
each term.  Compared to standard classical force fields, the inclusion
of explicit polarization and many-body dispersion leads to larger
evaluation times: $1-100$~s for systems composed of $10-100$ atoms on
a single core, respectively.  Notably, roughly 90\% of this time is
spent predicting the multipoles, due to the large training set and
complexity of the aSLATM representation.  While such an evaluation
time is significant, several strategies may be devised in the context
of a molecular dynamics simulation.  For instance, multipoles may
remain frozen and only get updated when large conformational changes
are detected.

We presented electrostatic calculations using distributed
multipole---up to quadrupole---models.  In comparison with other
atomic properties, an accurate prediction of multipole electrostatics
proves all the more challenging, and critical for the accurate
estimation of various molecular systems.  Improvements will require
more accurate models, and possibly the incorporation of more advanced
physical interactions, such as anisotropic\cite{van2017new} or
many-body repulsion interactions.  Our framework paves the way toward
significantly more transferable models that blend in the physical laws
and symmetries relevant for the phenomena at hand with a data-driven
approach to infer the variation of environmentally-dependent local
atomic parameters across chemical space.  We expect such models that
are transferable across chemical composition to be of use in systems
of interest in chemistry, biology, and materials science.

\section*{Acknowledgments}


We thank Denis Andrienko, Omar Valsson, and Alessandro de Vita for
critical discussions and Lori A.~Burns and C.~David Sherrill for
access to the SSI database.

T.B.~acknowledges funding from an Emmy Noether Fellowship of the
German Research Foundation (DFG). R.D.~acknowledges partial support
from Cornell University through startup funding and the Cornell Center
for Materials Research with funding from the NSF MRSEC program
(DMR-1719875). A.T.~acknowledges funding from the European Research
Council (ERC Consolidator Grant BeStMo).  O.A.v.L.~acknowledges
funding from the Swiss National Science foundation (No.~PP00P2\_138932
and 407540\_167186 NFP 75 Big Data).  This research was partly
supported by the NCCR MARVEL, funded by the Swiss National Science
Foundation.

\appendix
\section{Many-body dispersion}
\label{app:mbd}

The following summarizes the many-body dispersion (MBD)
interaction\cite{tkatchenko2009accurate, tkatchenko2012accurate,
  hermann2017first} as implemented elsewhere.\cite{bereau2014toward}
We recall the atomic polarizability $\alpha_p$ of atom $p$.  The
frequency dependence of $\alpha_p$ allows for an estimation of the
pairwise dispersion coefficient via the Casimir-Polder integral
\begin{equation}
  C_{6pq} = \frac 3\pi \int_0^\infty {\rm d}\omega \alpha_p(i\omega)
  \alpha_q(i\omega),
\end{equation}
where $i\omega$ are imaginary frequencies and $p$ and $q$ are a pair
of atoms.  Given reference free-atom values for $C_{6pp}$, we can
estimate the characteristic frequency of atom $p$ $\omega_p = 4
C_{6pp}/3\alpha_p^2$.\cite{chu2004linear}

The atomic polarizabilities and characteristic frequencies yield the
necessary ingredients for the system of coupled quantum harmonic
oscillators with $N$ atoms
\begin{equation}
  C_{pq}^{\rm QHO} = \omega_p^2 \delta_{pq} + (1-\delta_{pq})\omega_p
  \omega_q \sqrt{\alpha_p \alpha_q} \mathcal{T}_{pq},
\end{equation}
where $\mathcal{T}_{pq} = \nabla_{{\bf r}_p} \otimes \nabla_{{\bf
    r}_q} W(r_{pq})$ is a dipole interaction tensor with modified
Coulomb potential
\begin{equation}
  W(r_{pq}) = \frac{1-\exp{\left[- \left(\frac{r_{pq}}{R_{pq}^{\rm
            vdW}}\right)^\beta\right]}}{r_{pq}}.
\end{equation}
In this equation, $\beta$ is a range-separation parameter and
$R_{pq}^{\rm vdW} = \gamma (R_p^{\rm vdW} + R_q^{\rm vdW})$ is the sum
of effective van der Waals radii scaled by a chemistry-independent
fitting parameter.  The effective van der Waals radius is obtained by
scaling its reference free-atom counterpart: $R_p^{\rm vdW} =
(\alpha_p/\alpha_p^{\rm free})^{1/3} R_p^{\rm vdW, free}$.  An
expression for $\mathcal{T}_{pq}$ is provided in Bereau \emph{et
  al.}\cite{bereau2014toward} In particular, we apply a range
separation to the dipole interaction tensor by scaling it by a Fermi
function\cite{ambrosetti2014long}
\begin{equation}
  f(r_{pq}) = \frac 1{1+\exp{\left[ - d (r_{pq}/R_{pq}^{\rm vdW} -
        1)\right]}}.
\end{equation}

Diagonalizing the $3N \times 3N$ matrix $C_{pq}^{\rm QHO}$ yields its
eigenvalues $\{ \lambda_i \}$, which in turn provide the MBD energy
\begin{equation}
  E_{\rm MBD} = \frac 12 \sum_{i=1}^{3N} \sqrt{\lambda_i} - \frac 32
  \sum_{p=1}^N \omega_p.
\end{equation}
The methodology depends on three chemistry-independent parameters:
$\beta$, $\gamma$, and $d$.

\section{Covariant kernels}
\label{app:mtp_cov}

Glielmo et al.\cite{glielmo2017accurate}~recently proposed a covariant kernel
${\bf K}^\mu$ for vector quantities---suitable here to predict dipoles---such
that two samples $\rho$ and $\rho'$ subject to rotations $\mathcal{S}$ and
$\mathcal{S}'$, respectively, will obey
\begin{equation}
  {\bf K}^\mu(\mathcal{S}\rho, \mathcal{S}'\rho') = 
  {\bf SK}^\mu(\rho, \rho'){\bf S}'^{\rm T}.
\end{equation}
The atom $i$ from sample $\rho$ is encoded by a set of atom-centered Gaussian
functions
\begin{equation}
  \rho({\bf r}, \{ {\bf r}_i \}) = \frac{1}{(2\pi \sigma^2)^{3/2}} 
  \sum_i \exp\left( -\frac{||{\bf r} - {\bf r}_i||^2}{2\sigma^2} \right),
\end{equation}
and the covariant kernel is analytically integrated over all 3D rotations to
yield\cite{glielmo2017accurate}
\begin{align}
  {\bf K}^\mu(\rho, \rho') = \frac 1{L} \sum_{ij}
  \phi(r_i, r_j) {\bf r}_i \otimes {\bf r}_j'^{\rm T}, \\
  \phi(r_i, r_j) = \frac{\exp\left(-\alpha_{ij}\right)}{\gamma_{ij}^2} 
  \left( \gamma_{ij}\cosh \gamma_{ij} - \sinh \gamma_{ij}\right), \notag \\
  L = (2\sqrt{\pi\sigma^2})^3, \quad
  \alpha_{ij} = \frac{r_i^2 + r_j^2}{4\sigma^2}, \quad 
  \gamma_{ij} = \frac{r_ir_j}{2\sigma^2}, \notag
\end{align}
where $\otimes$ denotes the outer product.

In the present work, we extend the construction of covariant kernels to
predict quadrupole moments.  Following a similar procedure adapted to
second-rank tensors, we enforce the relation
\begin{equation}
  {\bf K}^{\rm Q}(\mathcal{S}\rho, \mathcal{S}'\rho') =
  {\bf S}'{\bf S}^{\rm T}{\bf K}^{\rm Q}(\rho, \rho'){\bf S}{\bf S}'^{\rm T}
\end{equation}
onto a base pairwise kernel of diagonal form: ${\bf K}^{\rm b}(\rho, \rho') =
\mathbb{1} k^{\rm b}(\rho, \rho')$, where $k^{\rm b}(\rho, \rho')$ is
independent of the reference frame.  The covariant kernel is constructed by
integrating the base kernel over all 3D rotations
\begin{equation}
  {\bf K}^{\rm Q}(\rho, \rho') = \frac 1L \sum_{ij} \int {\rm d}\mathcal{S}
  {\bf S} \otimes {\bf S}^{\rm T} k^{\rm b}(\rho, \mathcal{S}^{\rm T} \rho'),
\end{equation}
which leads to the expression
\begin{align}
  \label{eq:appquadall}
  {\bf K}^{\rm Q}(\rho, \rho') = \frac 1L \sum_{ij} 
  \left({\bf R}_j^{\rm T} \otimes {\bf R}_i\right) \Phi(r_i, r_j) 
  \left({{\bf R}_i}^{\rm T} \otimes {\bf R}_j\right), \notag\\
  \Phi(r_i, r_j) = \int {\rm d} \mathcal{\tilde R} \tilde{\bf R}^{\rm T}
  \otimes \tilde{\bf R} k^{\rm b}(\tilde {\bf r}_i, \tilde {\bf R}\tilde {\bf
    r}'_j) ,
\end{align}
where ${\bf R}_i$ and ${\bf R}_j$ are the rotation matrices that align ${\bf
  r}_i$ and ${\bf r}_j$ onto the $z$ axis to form $\tilde {\bf r}_i$ and
$\tilde {\bf r}'_j$, respectively.\cite{glielmo2017accurate}  We analytically
integrate all 3D rotations
\begin{align}
  \label{eq:appquad}
  \Phi(r_i, r_j) =& {\rm e}^{\frac{-\alpha_{ij}^2}{4\sigma^2}} \int {\rm
      d}\alpha \int {\rm d}\beta \int {\rm d}\gamma \frac{\sin \beta}{8\pi^2} \notag\\
  & \times {\bf R}^{\rm
    T}(\alpha,\beta,\gamma) \otimes {\bf R}(\alpha,\beta,\gamma)
  {\rm e}^{\frac{r_ir_j\cos\beta}{2\sigma^2}} \notag\\
  =& \left(
\begin{array}{ccccccccc}
\varphi_1 &0&0&0&\varphi_2&0&0&0&0\\
0&\varphi_1&0&-\varphi_2&0&0&0&0&0\\
0&0&\varphi_3&0&0&0&0&0&0\\
0&-\varphi_2&0&\varphi_1&0&0&0&0&0\\
\varphi_2&0&0&0&\varphi_1&0&0&0&0\\
0&0&0&0&0&\varphi_3&0&0&0\\
0&0&0&0&0&0&\varphi_3&0&0\\
0&0&0&0&0&0&0&\varphi_3&0\\
0&0&0&0&0&0&0&0&\varphi_4\\
\end{array} 
\right),
\end{align}
where
\begin{align}
  \varphi_1 &= \frac{{\rm
      e}^{\frac{-\alpha_{ij}^2}{4\sigma^2}}}{4\gamma_{ij}^2} \left(
    \gamma^2_{ij} \sinh \gamma_{ij} - \gamma_{ij} \cosh \gamma_{ij} +
    \sinh\gamma_{ij} \right) \notag\\
  \varphi_2 &= \frac{{\rm e}^{\frac{-\alpha_{ij}^2}{4\sigma^2}}}{4\gamma_{ij}}
  \left( \gamma_{ij} \cosh \gamma_{ij} -
    \sinh\gamma_{ij} \right) \notag\\
  \varphi_3 &= \frac{{\rm e}^{\frac{-\alpha_{ij}^2}{4\sigma^2}}}{2\gamma_{ij}^2}
  \left( \gamma_{ij} \cosh \gamma_{ij} -
    \sinh\gamma_{ij} \right) \notag\\
  \varphi_4 &= \frac{{\rm e}^{\frac{-\alpha_{ij}^2}{4\sigma^2}}}{\gamma_{ij}^2}
  \left( \frac{\gamma^2_{ij}}2 \sinh \gamma_{ij} - \gamma_{ij} \cosh \gamma_{ij} +
    \sinh\gamma_{ij} \right). 
\end{align}

\bibliography{biblio}  

\begin{thebibliography}{72}%
\makeatletter
\providecommand \@ifxundefined [1]{%
 \@ifx{#1\undefined}
}%
\providecommand \@ifnum [1]{%
 \ifnum #1\expandafter \@firstoftwo
 \else \expandafter \@secondoftwo
 \fi
}%
\providecommand \@ifx [1]{%
 \ifx #1\expandafter \@firstoftwo
 \else \expandafter \@secondoftwo
 \fi
}%
\providecommand \natexlab [1]{#1}%
\providecommand \enquote  [1]{``#1''}%
\providecommand \bibnamefont  [1]{#1}%
\providecommand \bibfnamefont [1]{#1}%
\providecommand \citenamefont [1]{#1}%
\providecommand \href@noop [0]{\@secondoftwo}%
\providecommand \href [0]{\begingroup \@sanitize@url \@href}%
\providecommand \@href[1]{\@@startlink{#1}\@@href}%
\providecommand \@@href[1]{\endgroup#1\@@endlink}%
\providecommand \@sanitize@url [0]{\catcode `\\12\catcode `\$12\catcode
  `\&12\catcode `\#12\catcode `\^12\catcode `\_12\catcode `\%12\relax}%
\providecommand \@@startlink[1]{}%
\providecommand \@@endlink[0]{}%
\providecommand \url  [0]{\begingroup\@sanitize@url \@url }%
\providecommand \@url [1]{\endgroup\@href {#1}{\urlprefix }}%
\providecommand \urlprefix  [0]{URL }%
\providecommand \Eprint [0]{\href }%
\providecommand \doibase [0]{http://dx.doi.org/}%
\providecommand \selectlanguage [0]{\@gobble}%
\providecommand \bibinfo  [0]{\@secondoftwo}%
\providecommand \bibfield  [0]{\@secondoftwo}%
\providecommand \translation [1]{[#1]}%
\providecommand \BibitemOpen [0]{}%
\providecommand \bibitemStop [0]{}%
\providecommand \bibitemNoStop [0]{.\EOS\space}%
\providecommand \EOS [0]{\spacefactor3000\relax}%
\providecommand \BibitemShut  [1]{\csname bibitem#1\endcsname}%
\let\auto@bib@innerbib\@empty
\bibitem [{\citenamefont {Grimme}(2014)}]{grimme2014general}%
  \BibitemOpen
  \bibfield  {author} {\bibinfo {author} {\bibfnamefont {S.}~\bibnamefont
  {Grimme}},\ }\href@noop {} {\bibfield  {journal} {\bibinfo  {journal} {J.
  Chem. Theory Comput.}\ }\textbf {\bibinfo {volume} {10}},\ \bibinfo {pages}
  {4497} (\bibinfo {year} {2014})}\BibitemShut {NoStop}%
\bibitem [{\citenamefont {Metz}, \citenamefont {Piszczatowski},\ and\
  \citenamefont {Szalewicz}(2016)}]{metz2016automatic}%
  \BibitemOpen
  \bibfield  {author} {\bibinfo {author} {\bibfnamefont {M.~P.}\ \bibnamefont
  {Metz}}, \bibinfo {author} {\bibfnamefont {K.}~\bibnamefont {Piszczatowski}},
  \ and\ \bibinfo {author} {\bibfnamefont {K.}~\bibnamefont {Szalewicz}},\
  }\href@noop {} {\bibfield  {journal} {\bibinfo  {journal} {J. Chem. Theory
  Comput.}\ }\textbf {\bibinfo {volume} {12}},\ \bibinfo {pages} {5895}
  (\bibinfo {year} {2016})}\BibitemShut {NoStop}%
\bibitem [{\citenamefont {Jeziorski}, \citenamefont {Moszynski},\ and\
  \citenamefont {Szalewicz}(1994)}]{jeziorski1994perturbation}%
  \BibitemOpen
  \bibfield  {author} {\bibinfo {author} {\bibfnamefont {B.}~\bibnamefont
  {Jeziorski}}, \bibinfo {author} {\bibfnamefont {R.}~\bibnamefont
  {Moszynski}}, \ and\ \bibinfo {author} {\bibfnamefont {K.}~\bibnamefont
  {Szalewicz}},\ }\href@noop {} {\bibfield  {journal} {\bibinfo  {journal}
  {Chem.~Rev.}\ }\textbf {\bibinfo {volume} {94}},\ \bibinfo {pages} {1887}
  (\bibinfo {year} {1994})}\BibitemShut {NoStop}%
\bibitem [{\citenamefont {Van~Vleet}\ \emph {et~al.}(2016)\citenamefont
  {Van~Vleet}, \citenamefont {Misquitta}, \citenamefont {Stone},\ and\
  \citenamefont {Schmidt}}]{van2016beyond}%
  \BibitemOpen
  \bibfield  {author} {\bibinfo {author} {\bibfnamefont {M.~J.}\ \bibnamefont
  {Van~Vleet}}, \bibinfo {author} {\bibfnamefont {A.~J.}\ \bibnamefont
  {Misquitta}}, \bibinfo {author} {\bibfnamefont {A.~J.}\ \bibnamefont
  {Stone}}, \ and\ \bibinfo {author} {\bibfnamefont {J.}~\bibnamefont
  {Schmidt}},\ }\href@noop {} {\bibfield  {journal} {\bibinfo  {journal} {J.
  Chem. Theory Comput.}\ }\textbf {\bibinfo {volume} {12}},\ \bibinfo {pages}
  {3851} (\bibinfo {year} {2016})}\BibitemShut {NoStop}%
\bibitem [{\citenamefont {Vandenbrande}\ \emph {et~al.}(2017)\citenamefont
  {Vandenbrande}, \citenamefont {Waroquier}, \citenamefont {Speybroeck},\ and\
  \citenamefont {Verstraelen}}]{Vandenbrande2017}%
  \BibitemOpen
  \bibfield  {author} {\bibinfo {author} {\bibfnamefont {S.}~\bibnamefont
  {Vandenbrande}}, \bibinfo {author} {\bibfnamefont {M.}~\bibnamefont
  {Waroquier}}, \bibinfo {author} {\bibfnamefont {V.~V.}\ \bibnamefont
  {Speybroeck}}, \ and\ \bibinfo {author} {\bibfnamefont {T.}~\bibnamefont
  {Verstraelen}},\ }\href@noop {} {\bibfield  {journal} {\bibinfo  {journal}
  {J. Chem. Theory Comput.}\ }\textbf {\bibinfo {volume} {13}},\ \bibinfo
  {pages} {161} (\bibinfo {year} {2017})}\BibitemShut {NoStop}%
\bibitem [{\citenamefont {Cole}\ \emph {et~al.}(2016)\citenamefont {Cole},
  \citenamefont {Vilseck}, \citenamefont {Tirado-Rives}, \citenamefont
  {Payne},\ and\ \citenamefont {Jorgensen}}]{cole2016biomolecular}%
  \BibitemOpen
  \bibfield  {author} {\bibinfo {author} {\bibfnamefont {D.~J.}\ \bibnamefont
  {Cole}}, \bibinfo {author} {\bibfnamefont {J.~Z.}\ \bibnamefont {Vilseck}},
  \bibinfo {author} {\bibfnamefont {J.}~\bibnamefont {Tirado-Rives}}, \bibinfo
  {author} {\bibfnamefont {M.~C.}\ \bibnamefont {Payne}}, \ and\ \bibinfo
  {author} {\bibfnamefont {W.~L.}\ \bibnamefont {Jorgensen}},\ }\href@noop {}
  {\bibfield  {journal} {\bibinfo  {journal} {J. Chem. Theory Comput.}\
  }\textbf {\bibinfo {volume} {12}},\ \bibinfo {pages} {2312} (\bibinfo {year}
  {2016})}\BibitemShut {NoStop}%
\bibitem [{\citenamefont {Bart{\'o}k}\ \emph {et~al.}(2010)\citenamefont
  {Bart{\'o}k}, \citenamefont {Payne}, \citenamefont {Kondor},\ and\
  \citenamefont {Cs{\'a}nyi}}]{bartok2010gaussian}%
  \BibitemOpen
  \bibfield  {author} {\bibinfo {author} {\bibfnamefont {A.~P.}\ \bibnamefont
  {Bart{\'o}k}}, \bibinfo {author} {\bibfnamefont {M.~C.}\ \bibnamefont
  {Payne}}, \bibinfo {author} {\bibfnamefont {R.}~\bibnamefont {Kondor}}, \
  and\ \bibinfo {author} {\bibfnamefont {G.}~\bibnamefont {Cs{\'a}nyi}},\
  }\href@noop {} {\bibfield  {journal} {\bibinfo  {journal} {Phys. Rev. Lett.}\
  }\textbf {\bibinfo {volume} {104}},\ \bibinfo {pages} {136403} (\bibinfo
  {year} {2010})}\BibitemShut {NoStop}%
\bibitem [{\citenamefont {Li}, \citenamefont {Kermode},\ and\ \citenamefont
  {De~Vita}(2015)}]{li2015molecular}%
  \BibitemOpen
  \bibfield  {author} {\bibinfo {author} {\bibfnamefont {Z.}~\bibnamefont
  {Li}}, \bibinfo {author} {\bibfnamefont {J.~R.}\ \bibnamefont {Kermode}}, \
  and\ \bibinfo {author} {\bibfnamefont {A.}~\bibnamefont {De~Vita}},\
  }\href@noop {} {\bibfield  {journal} {\bibinfo  {journal} {Phys. Rev. Lett.}\
  }\textbf {\bibinfo {volume} {114}},\ \bibinfo {pages} {096405} (\bibinfo
  {year} {2015})}\BibitemShut {NoStop}%
\bibitem [{\citenamefont {Behler}(2016)}]{behler2016perspective}%
  \BibitemOpen
  \bibfield  {author} {\bibinfo {author} {\bibfnamefont {J.}~\bibnamefont
  {Behler}},\ }\href@noop {} {\bibfield  {journal} {\bibinfo  {journal} {J.
  Chem. Phys.}\ }\textbf {\bibinfo {volume} {145}},\ \bibinfo {pages} {170901}
  (\bibinfo {year} {2016})}\BibitemShut {NoStop}%
\bibitem [{\citenamefont {Chmiela}\ \emph {et~al.}(2017)\citenamefont
  {Chmiela}, \citenamefont {Tkatchenko}, \citenamefont {Sauceda}, \citenamefont
  {Poltavsky}, \citenamefont {Sch{\"u}tt},\ and\ \citenamefont
  {M{\"u}ller}}]{chmiela2017machine}%
  \BibitemOpen
  \bibfield  {author} {\bibinfo {author} {\bibfnamefont {S.}~\bibnamefont
  {Chmiela}}, \bibinfo {author} {\bibfnamefont {A.}~\bibnamefont {Tkatchenko}},
  \bibinfo {author} {\bibfnamefont {H.~E.}\ \bibnamefont {Sauceda}}, \bibinfo
  {author} {\bibfnamefont {I.}~\bibnamefont {Poltavsky}}, \bibinfo {author}
  {\bibfnamefont {K.~T.}\ \bibnamefont {Sch{\"u}tt}}, \ and\ \bibinfo {author}
  {\bibfnamefont {K.-R.}\ \bibnamefont {M{\"u}ller}},\ }\href@noop {}
  {\bibfield  {journal} {\bibinfo  {journal} {Sci.~Adv.}\ }\textbf {\bibinfo
  {volume} {3}},\ \bibinfo {pages} {e1603015} (\bibinfo {year}
  {2017})}\BibitemShut {NoStop}%
\bibitem [{\citenamefont {Botu}\ \emph {et~al.}(2016)\citenamefont {Botu},
  \citenamefont {Batra}, \citenamefont {Chapman},\ and\ \citenamefont
  {Ramprasad}}]{botu2016machine}%
  \BibitemOpen
  \bibfield  {author} {\bibinfo {author} {\bibfnamefont {V.}~\bibnamefont
  {Botu}}, \bibinfo {author} {\bibfnamefont {R.}~\bibnamefont {Batra}},
  \bibinfo {author} {\bibfnamefont {J.}~\bibnamefont {Chapman}}, \ and\
  \bibinfo {author} {\bibfnamefont {R.}~\bibnamefont {Ramprasad}},\ }\href@noop
  {} {\bibfield  {journal} {\bibinfo  {journal} {J.~Phys.~Chem.~C}\ }\textbf
  {\bibinfo {volume} {121}},\ \bibinfo {pages} {511} (\bibinfo {year}
  {2016})}\BibitemShut {NoStop}%
\bibitem [{\citenamefont {Sch{\"u}tt}\ \emph {et~al.}(2017)\citenamefont
  {Sch{\"u}tt}, \citenamefont {Arbabzadah}, \citenamefont {Chmiela},
  \citenamefont {M{\"u}ller},\ and\ \citenamefont
  {Tkatchenko}}]{schutt2017quantum}%
  \BibitemOpen
  \bibfield  {author} {\bibinfo {author} {\bibfnamefont {K.~T.}\ \bibnamefont
  {Sch{\"u}tt}}, \bibinfo {author} {\bibfnamefont {F.}~\bibnamefont
  {Arbabzadah}}, \bibinfo {author} {\bibfnamefont {S.}~\bibnamefont {Chmiela}},
  \bibinfo {author} {\bibfnamefont {K.~R.}\ \bibnamefont {M{\"u}ller}}, \ and\
  \bibinfo {author} {\bibfnamefont {A.}~\bibnamefont {Tkatchenko}},\
  }\href@noop {} {\bibfield  {journal} {\bibinfo  {journal} {Nat.~Comm.}\
  }\textbf {\bibinfo {volume} {8}},\ \bibinfo {pages} {13890} (\bibinfo {year}
  {2017})}\BibitemShut {NoStop}%
\bibitem [{\citenamefont {Natarajan}, \citenamefont {Morawietz},\ and\
  \citenamefont {Behler}(2015)}]{natarajan2015representing}%
  \BibitemOpen
  \bibfield  {author} {\bibinfo {author} {\bibfnamefont {S.~K.}\ \bibnamefont
  {Natarajan}}, \bibinfo {author} {\bibfnamefont {T.}~\bibnamefont
  {Morawietz}}, \ and\ \bibinfo {author} {\bibfnamefont {J.}~\bibnamefont
  {Behler}},\ }\href@noop {} {\bibfield  {journal} {\bibinfo  {journal} {Phys.
  Chem. Chem. Phys.}\ }\textbf {\bibinfo {volume} {17}},\ \bibinfo {pages}
  {8356} (\bibinfo {year} {2015})}\BibitemShut {NoStop}%
\bibitem [{\citenamefont {Rupp}\ \emph {et~al.}(2012)\citenamefont {Rupp},
  \citenamefont {Tkatchenko}, \citenamefont {M{\"u}ller},\ and\ \citenamefont
  {Von~Lilienfeld}}]{rupp2012fast}%
  \BibitemOpen
  \bibfield  {author} {\bibinfo {author} {\bibfnamefont {M.}~\bibnamefont
  {Rupp}}, \bibinfo {author} {\bibfnamefont {A.}~\bibnamefont {Tkatchenko}},
  \bibinfo {author} {\bibfnamefont {K.-R.}\ \bibnamefont {M{\"u}ller}}, \ and\
  \bibinfo {author} {\bibfnamefont {O.~A.}\ \bibnamefont {Von~Lilienfeld}},\
  }\href@noop {} {\bibfield  {journal} {\bibinfo  {journal} {Phys. Rev. Lett.}\
  }\textbf {\bibinfo {volume} {108}},\ \bibinfo {pages} {058301} (\bibinfo
  {year} {2012})}\BibitemShut {NoStop}%
\bibitem [{\citenamefont {Hansen}\ \emph {et~al.}(2013)\citenamefont {Hansen},
  \citenamefont {Montavon}, \citenamefont {Biegler}, \citenamefont {Fazli},
  \citenamefont {Rupp}, \citenamefont {Scheffler}, \citenamefont
  {Von~Lilienfeld}, \citenamefont {Tkatchenko},\ and\ \citenamefont
  {Müller}}]{hansen2013assessment}%
  \BibitemOpen
  \bibfield  {author} {\bibinfo {author} {\bibfnamefont {K.}~\bibnamefont
  {Hansen}}, \bibinfo {author} {\bibfnamefont {G.}~\bibnamefont {Montavon}},
  \bibinfo {author} {\bibfnamefont {F.}~\bibnamefont {Biegler}}, \bibinfo
  {author} {\bibfnamefont {S.}~\bibnamefont {Fazli}}, \bibinfo {author}
  {\bibfnamefont {M.}~\bibnamefont {Rupp}}, \bibinfo {author} {\bibfnamefont
  {M.}~\bibnamefont {Scheffler}}, \bibinfo {author} {\bibfnamefont {O.~A.}\
  \bibnamefont {Von~Lilienfeld}}, \bibinfo {author} {\bibfnamefont
  {A.}~\bibnamefont {Tkatchenko}}, \ and\ \bibinfo {author} {\bibfnamefont
  {K.-R.}\ \bibnamefont {Müller}},\ }\href@noop {} {\bibfield  {journal}
  {\bibinfo  {journal} {J. Chem. Theory Comput.}\ }\textbf {\bibinfo {volume}
  {9}},\ \bibinfo {pages} {3404} (\bibinfo {year} {2013})}\BibitemShut
  {NoStop}%
\bibitem [{\citenamefont {Ramakrishnan}\ and\ \citenamefont {von
  Lilienfeld}(2017)}]{ramakrishnan2017machine}%
  \BibitemOpen
  \bibfield  {author} {\bibinfo {author} {\bibfnamefont {R.}~\bibnamefont
  {Ramakrishnan}}\ and\ \bibinfo {author} {\bibfnamefont {O.~A.}\ \bibnamefont
  {von Lilienfeld}},\ }\enquote {\bibinfo {title} {Machine learning, quantum
  chemistry, and chemical space},}\ in\ \href@noop {} {\emph {\bibinfo
  {booktitle} {Rev.~Comp.~Ch.}}}\ (\bibinfo  {publisher} {John Wiley \& Sons,
  Inc.},\ \bibinfo {year} {2017})\ pp.\ \bibinfo {pages} {225--256}\BibitemShut
  {NoStop}%
\bibitem [{\citenamefont {Bereau}\ and\ \citenamefont
  {Kremer}(2015)}]{bereau2015automated}%
  \BibitemOpen
  \bibfield  {author} {\bibinfo {author} {\bibfnamefont {T.}~\bibnamefont
  {Bereau}}\ and\ \bibinfo {author} {\bibfnamefont {K.}~\bibnamefont
  {Kremer}},\ }\href@noop {} {\bibfield  {journal} {\bibinfo  {journal} {J.
  Chem. Theory Comput.}\ }\textbf {\bibinfo {volume} {11}},\ \bibinfo {pages}
  {2783} (\bibinfo {year} {2015})}\BibitemShut {NoStop}%
\bibitem [{\citenamefont {Bereau}, \citenamefont {Andrienko},\ and\
  \citenamefont {von Lilienfeld}(2015)}]{bereau2015transferable}%
  \BibitemOpen
  \bibfield  {author} {\bibinfo {author} {\bibfnamefont {T.}~\bibnamefont
  {Bereau}}, \bibinfo {author} {\bibfnamefont {D.}~\bibnamefont {Andrienko}}, \
  and\ \bibinfo {author} {\bibfnamefont {O.~A.}\ \bibnamefont {von
  Lilienfeld}},\ }\href@noop {} {\bibfield  {journal} {\bibinfo  {journal} {J.
  Chem. Theory Comput.}\ }\textbf {\bibinfo {volume} {11}},\ \bibinfo {pages}
  {3225} (\bibinfo {year} {2015})}\BibitemShut {NoStop}%
\bibitem [{\citenamefont {Stone}(2013)}]{stone2013theory}%
  \BibitemOpen
  \bibfield  {author} {\bibinfo {author} {\bibfnamefont {A.}~\bibnamefont
  {Stone}},\ }\href@noop {} {\emph {\bibinfo {title} {The theory of
  intermolecular forces}}}\ (\bibinfo  {publisher} {Oxford University Press},\
  \bibinfo {year} {2013})\BibitemShut {NoStop}%
\bibitem [{\citenamefont {Jure{\v{c}}ka}\ \emph {et~al.}(2006)\citenamefont
  {Jure{\v{c}}ka}, \citenamefont {{\v{S}}poner}, \citenamefont
  {{\v{C}}ern{\`y}},\ and\ \citenamefont {Hobza}}]{jurevcka2006benchmark}%
  \BibitemOpen
  \bibfield  {author} {\bibinfo {author} {\bibfnamefont {P.}~\bibnamefont
  {Jure{\v{c}}ka}}, \bibinfo {author} {\bibfnamefont {J.}~\bibnamefont
  {{\v{S}}poner}}, \bibinfo {author} {\bibfnamefont {J.}~\bibnamefont
  {{\v{C}}ern{\`y}}}, \ and\ \bibinfo {author} {\bibfnamefont {P.}~\bibnamefont
  {Hobza}},\ }\href@noop {} {\bibfield  {journal} {\bibinfo  {journal} {Phys.
  Chem. Chem. Phys.}\ }\textbf {\bibinfo {volume} {8}},\ \bibinfo {pages}
  {1985} (\bibinfo {year} {2006})}\BibitemShut {NoStop}%
\bibitem [{\citenamefont {Zhao}\ and\ \citenamefont
  {Truhlar}(2008)}]{zhao2008m06}%
  \BibitemOpen
  \bibfield  {author} {\bibinfo {author} {\bibfnamefont {Y.}~\bibnamefont
  {Zhao}}\ and\ \bibinfo {author} {\bibfnamefont {D.~G.}\ \bibnamefont
  {Truhlar}},\ }\href@noop {} {\bibfield  {journal} {\bibinfo  {journal}
  {Theor.~Chem.~Acc.}\ }\textbf {\bibinfo {volume} {120}},\ \bibinfo {pages}
  {215} (\bibinfo {year} {2008})}\BibitemShut {NoStop}%
\bibitem [{\citenamefont {Misquitta}, \citenamefont {Stone},\ and\
  \citenamefont {Fazeli}(2014)}]{misquitta2014distributed}%
  \BibitemOpen
  \bibfield  {author} {\bibinfo {author} {\bibfnamefont {A.~J.}\ \bibnamefont
  {Misquitta}}, \bibinfo {author} {\bibfnamefont {A.~J.}\ \bibnamefont
  {Stone}}, \ and\ \bibinfo {author} {\bibfnamefont {F.}~\bibnamefont
  {Fazeli}},\ }\href@noop {} {\bibfield  {journal} {\bibinfo  {journal} {J.
  Chem. Theory Comput.}\ }\textbf {\bibinfo {volume} {10}},\ \bibinfo {pages}
  {5405} (\bibinfo {year} {2014})}\BibitemShut {NoStop}%
\bibitem [{\citenamefont {Verstraelen}\ \emph {et~al.}()\citenamefont
  {Verstraelen}, \citenamefont {Tecmer}, \citenamefont {Heidar-Zadeh},
  \citenamefont {Boguslawski}, \citenamefont {Chan}, \citenamefont {Zhao},
  \citenamefont {Kim}, \citenamefont {Vandenbrande}, \citenamefont {Yang},
  \citenamefont {Gonz\'alez-Espinoza}, \citenamefont {Fias}, \citenamefont
  {Limacher}, \citenamefont {Berrocal}, \citenamefont {Malek},\ and\
  \citenamefont {Ayers}}]{horton}%
  \BibitemOpen
  \bibfield  {author} {\bibinfo {author} {\bibfnamefont {T.}~\bibnamefont
  {Verstraelen}}, \bibinfo {author} {\bibfnamefont {P.}~\bibnamefont {Tecmer}},
  \bibinfo {author} {\bibfnamefont {F.}~\bibnamefont {Heidar-Zadeh}}, \bibinfo
  {author} {\bibfnamefont {K.}~\bibnamefont {Boguslawski}}, \bibinfo {author}
  {\bibfnamefont {M.}~\bibnamefont {Chan}}, \bibinfo {author} {\bibfnamefont
  {Y.}~\bibnamefont {Zhao}}, \bibinfo {author} {\bibfnamefont {T.~D.}\
  \bibnamefont {Kim}}, \bibinfo {author} {\bibfnamefont {S.}~\bibnamefont
  {Vandenbrande}}, \bibinfo {author} {\bibfnamefont {D.}~\bibnamefont {Yang}},
  \bibinfo {author} {\bibfnamefont {C.~E.}\ \bibnamefont
  {Gonz\'alez-Espinoza}}, \bibinfo {author} {\bibfnamefont {S.}~\bibnamefont
  {Fias}}, \bibinfo {author} {\bibfnamefont {P.~A.}\ \bibnamefont {Limacher}},
  \bibinfo {author} {\bibfnamefont {D.}~\bibnamefont {Berrocal}}, \bibinfo
  {author} {\bibfnamefont {A.}~\bibnamefont {Malek}}, \ and\ \bibinfo {author}
  {\bibfnamefont {P.~W.}\ \bibnamefont {Ayers}},\ }\href@noop {} {\enquote
  {\bibinfo {title} {{HORTON}, version 2.0.1},}\ }\bibinfo {howpublished}
  {\url{http://theochem.github.com/horton/}},\ \bibinfo {note} {accessed:
  2016-08-01}\BibitemShut {NoStop}%
\bibitem [{\citenamefont {Huang}\ and\ \citenamefont {von
  Lilienfeld}(2016)}]{huang2016communication}%
  \BibitemOpen
  \bibfield  {author} {\bibinfo {author} {\bibfnamefont {B.}~\bibnamefont
  {Huang}}\ and\ \bibinfo {author} {\bibfnamefont {O.~A.}\ \bibnamefont {von
  Lilienfeld}},\ }\href@noop {} {\bibfield  {journal} {\bibinfo  {journal}
  {J.~Chem.~Phys.}\ }\textbf {\bibinfo {volume} {145}},\ \bibinfo {pages}
  {161102} (\bibinfo {year} {2016})}\BibitemShut {NoStop}%
\bibitem [{\citenamefont {Huang}\ and\ \citenamefont {von
  Lilienfeld}(2017)}]{huang2017chemical}%
  \BibitemOpen
  \bibfield  {author} {\bibinfo {author} {\bibfnamefont {B.}~\bibnamefont
  {Huang}}\ and\ \bibinfo {author} {\bibfnamefont {O.~A.}\ \bibnamefont {von
  Lilienfeld}},\ }\href@noop {} {\bibfield  {journal} {\bibinfo  {journal}
  {arXiv preprint arXiv:1707.04146}\ } (\bibinfo {year} {2017})}\BibitemShut
  {NoStop}%
\bibitem [{\citenamefont {Christensen}\ \emph {et~al.}()\citenamefont
  {Christensen}, \citenamefont {Faber}, \citenamefont {Huang}, \citenamefont
  {Bratholm}, \citenamefont {Tkatchenko}, \citenamefont {M\"uller},\ and\
  \citenamefont {von Lilienfeld}}]{qml}%
  \BibitemOpen
  \bibfield  {author} {\bibinfo {author} {\bibfnamefont {A.~S.}\ \bibnamefont
  {Christensen}}, \bibinfo {author} {\bibfnamefont {F.~A.}\ \bibnamefont
  {Faber}}, \bibinfo {author} {\bibfnamefont {B.}~\bibnamefont {Huang}},
  \bibinfo {author} {\bibfnamefont {L.~A.}\ \bibnamefont {Bratholm}}, \bibinfo
  {author} {\bibfnamefont {A.}~\bibnamefont {Tkatchenko}}, \bibinfo {author}
  {\bibfnamefont {K.~R.}\ \bibnamefont {M\"uller}}, \ and\ \bibinfo {author}
  {\bibfnamefont {O.~A.}\ \bibnamefont {von Lilienfeld}},\ }\href@noop {}
  {\enquote {\bibinfo {title} {{QML: A Python Toolkit for Quantum Machine
  Learning}},}\ }\bibinfo {howpublished}
  {\url{https://github.com/qmlcode/qml}},\ \bibinfo {note} {accessed:
  2017-07-01}\BibitemShut {NoStop}%
\bibitem [{\citenamefont {Kim}, \citenamefont {Kim},\ and\ \citenamefont
  {Lee}(1981)}]{kim1981dependence}%
  \BibitemOpen
  \bibfield  {author} {\bibinfo {author} {\bibfnamefont {Y.~S.}\ \bibnamefont
  {Kim}}, \bibinfo {author} {\bibfnamefont {S.~K.}\ \bibnamefont {Kim}}, \ and\
  \bibinfo {author} {\bibfnamefont {W.~D.}\ \bibnamefont {Lee}},\ }\href@noop
  {} {\bibfield  {journal} {\bibinfo  {journal} {Chem. Phys. Lett.}\ }\textbf
  {\bibinfo {volume} {80}},\ \bibinfo {pages} {574} (\bibinfo {year}
  {1981})}\BibitemShut {NoStop}%
\bibitem [{\citenamefont {Lillestolen}\ and\ \citenamefont
  {Wheatley}(2008)}]{lillestolen08redefining}%
  \BibitemOpen
  \bibfield  {author} {\bibinfo {author} {\bibfnamefont {T.~C.}\ \bibnamefont
  {Lillestolen}}\ and\ \bibinfo {author} {\bibfnamefont {R.~J.}\ \bibnamefont
  {Wheatley}},\ }\href@noop {} {\bibfield  {journal} {\bibinfo  {journal}
  {Chem. Commun.}\ }\textbf {\bibinfo {volume} {45}},\ \bibinfo {pages} {5909}
  (\bibinfo {year} {2008})}\BibitemShut {NoStop}%
\bibitem [{\citenamefont {Verstraelen}\ \emph {et~al.}(2016)\citenamefont
  {Verstraelen}, \citenamefont {Vandenbrande}, \citenamefont {Heidar-Zadeh},
  \citenamefont {Vanduyfhuys}, \citenamefont {Van~Speybroeck}, \citenamefont
  {Waroquier},\ and\ \citenamefont {Ayers}}]{verstraelen2016minimal}%
  \BibitemOpen
  \bibfield  {author} {\bibinfo {author} {\bibfnamefont {T.}~\bibnamefont
  {Verstraelen}}, \bibinfo {author} {\bibfnamefont {S.}~\bibnamefont
  {Vandenbrande}}, \bibinfo {author} {\bibfnamefont {F.}~\bibnamefont
  {Heidar-Zadeh}}, \bibinfo {author} {\bibfnamefont {L.}~\bibnamefont
  {Vanduyfhuys}}, \bibinfo {author} {\bibfnamefont {V.}~\bibnamefont
  {Van~Speybroeck}}, \bibinfo {author} {\bibfnamefont {M.}~\bibnamefont
  {Waroquier}}, \ and\ \bibinfo {author} {\bibfnamefont {P.~W.}\ \bibnamefont
  {Ayers}},\ }\href@noop {} {\bibfield  {journal} {\bibinfo  {journal} {J.
  Chem. Theory Comput.}\ }\textbf {\bibinfo {volume} {12}},\ \bibinfo {pages}
  {3894} (\bibinfo {year} {2016})}\BibitemShut {NoStop}%
\bibitem [{\citenamefont {Hirshfeld}(1977)}]{hirshfeld1977bonded}%
  \BibitemOpen
  \bibfield  {author} {\bibinfo {author} {\bibfnamefont {F.~L.}\ \bibnamefont
  {Hirshfeld}},\ }\href@noop {} {\bibfield  {journal} {\bibinfo  {journal}
  {Theor.~Chim.~Acta}\ }\textbf {\bibinfo {volume} {44}},\ \bibinfo {pages}
  {129} (\bibinfo {year} {1977})}\BibitemShut {NoStop}%
\bibitem [{\citenamefont {Tkatchenko}\ and\ \citenamefont
  {Scheffler}(2009)}]{tkatchenko2009accurate}%
  \BibitemOpen
  \bibfield  {author} {\bibinfo {author} {\bibfnamefont {A.}~\bibnamefont
  {Tkatchenko}}\ and\ \bibinfo {author} {\bibfnamefont {M.}~\bibnamefont
  {Scheffler}},\ }\href@noop {} {\bibfield  {journal} {\bibinfo  {journal}
  {Phys. Rev. Lett.}\ }\textbf {\bibinfo {volume} {102}},\ \bibinfo {pages}
  {073005} (\bibinfo {year} {2009})}\BibitemShut {NoStop}%
\bibitem [{\citenamefont {Bereau}\ and\ \citenamefont {von
  Lilienfeld}(2014)}]{bereau2014toward}%
  \BibitemOpen
  \bibfield  {author} {\bibinfo {author} {\bibfnamefont {T.}~\bibnamefont
  {Bereau}}\ and\ \bibinfo {author} {\bibfnamefont {O.~A.}\ \bibnamefont {von
  Lilienfeld}},\ }\href@noop {} {\bibfield  {journal} {\bibinfo  {journal} {J.
  Chem. Phys.}\ }\textbf {\bibinfo {volume} {141}},\ \bibinfo {pages} {034101}
  (\bibinfo {year} {2014})}\BibitemShut {NoStop}%
\bibitem [{\citenamefont {Bu{\v{c}}ko}\ \emph {et~al.}(2014)\citenamefont
  {Bu{\v{c}}ko}, \citenamefont {Leb{\`e}gue}, \citenamefont {{\'A}ngy{\'a}n},\
  and\ \citenamefont {Hafner}}]{buvcko2014extending}%
  \BibitemOpen
  \bibfield  {author} {\bibinfo {author} {\bibfnamefont {T.}~\bibnamefont
  {Bu{\v{c}}ko}}, \bibinfo {author} {\bibfnamefont {S.}~\bibnamefont
  {Leb{\`e}gue}}, \bibinfo {author} {\bibfnamefont {J.~G.}\ \bibnamefont
  {{\'A}ngy{\'a}n}}, \ and\ \bibinfo {author} {\bibfnamefont {J.}~\bibnamefont
  {Hafner}},\ }\href@noop {} {\bibfield  {journal} {\bibinfo  {journal} {J.
  Chem. Phys.}\ }\textbf {\bibinfo {volume} {141}},\ \bibinfo {pages} {034114}
  (\bibinfo {year} {2014})}\BibitemShut {NoStop}%
\bibitem [{\citenamefont {Gobre}(2016)}]{gobre2016efficient}%
  \BibitemOpen
  \bibfield  {author} {\bibinfo {author} {\bibfnamefont {V.~V.}\ \bibnamefont
  {Gobre}},\ }\emph {\bibinfo {title} {Efficient modelling of linear electronic
  polarization in materials using atomic response functions}},\ \href@noop {}
  {Ph.D. thesis},\ \bibinfo  {school} {Technische Universit{\"a}t Berlin}
  (\bibinfo {year} {2016})\BibitemShut {NoStop}%
\bibitem [{\citenamefont {Adamo}\ and\ \citenamefont
  {Barone}(1999)}]{adamo1999toward}%
  \BibitemOpen
  \bibfield  {author} {\bibinfo {author} {\bibfnamefont {C.}~\bibnamefont
  {Adamo}}\ and\ \bibinfo {author} {\bibfnamefont {V.}~\bibnamefont {Barone}},\
  }\href@noop {} {\bibfield  {journal} {\bibinfo  {journal} {J. Chem. Phys.}\
  }\textbf {\bibinfo {volume} {110}},\ \bibinfo {pages} {6158} (\bibinfo {year}
  {1999})}\BibitemShut {NoStop}%
\bibitem [{\citenamefont {Kannemann}\ and\ \citenamefont
  {Becke}(2010)}]{kannemann2010van}%
  \BibitemOpen
  \bibfield  {author} {\bibinfo {author} {\bibfnamefont {F.~O.}\ \bibnamefont
  {Kannemann}}\ and\ \bibinfo {author} {\bibfnamefont {A.~D.}\ \bibnamefont
  {Becke}},\ }\href@noop {} {\bibfield  {journal} {\bibinfo  {journal} {J.
  Chem. Theory Comput.}\ }\textbf {\bibinfo {volume} {6}},\ \bibinfo {pages}
  {1081} (\bibinfo {year} {2010})}\BibitemShut {NoStop}%
\bibitem [{\citenamefont {Otero-de-la Roza}\ and\ \citenamefont
  {Johnson}(2013)}]{otero2013many}%
  \BibitemOpen
  \bibfield  {author} {\bibinfo {author} {\bibfnamefont {A.}~\bibnamefont
  {Otero-de-la Roza}}\ and\ \bibinfo {author} {\bibfnamefont {E.~R.}\
  \bibnamefont {Johnson}},\ }\href@noop {} {\bibfield  {journal} {\bibinfo
  {journal} {J. Chem. Phys.}\ }\textbf {\bibinfo {volume} {138}},\ \bibinfo
  {pages} {054103} (\bibinfo {year} {2013})}\BibitemShut {NoStop}%
\bibitem [{\citenamefont {Ren}\ and\ \citenamefont
  {Ponder}(2003)}]{ren2003polarizable}%
  \BibitemOpen
  \bibfield  {author} {\bibinfo {author} {\bibfnamefont {P.}~\bibnamefont
  {Ren}}\ and\ \bibinfo {author} {\bibfnamefont {J.~W.}\ \bibnamefont
  {Ponder}},\ }\href@noop {} {\bibfield  {journal} {\bibinfo  {journal} {J.
  Phys. Chem. B}\ }\textbf {\bibinfo {volume} {107}},\ \bibinfo {pages} {5933}
  (\bibinfo {year} {2003})}\BibitemShut {NoStop}%
\bibitem [{\citenamefont {Bereau}, \citenamefont {Kramer},\ and\ \citenamefont
  {Meuwly}(2013)}]{bereau2013leveraging}%
  \BibitemOpen
  \bibfield  {author} {\bibinfo {author} {\bibfnamefont {T.}~\bibnamefont
  {Bereau}}, \bibinfo {author} {\bibfnamefont {C.}~\bibnamefont {Kramer}}, \
  and\ \bibinfo {author} {\bibfnamefont {M.}~\bibnamefont {Meuwly}},\
  }\href@noop {} {\bibfield  {journal} {\bibinfo  {journal} {J. Chem. Theory
  Comput.}\ }\textbf {\bibinfo {volume} {9}},\ \bibinfo {pages} {5450}
  (\bibinfo {year} {2013})}\BibitemShut {NoStop}%
\bibitem [{\citenamefont {Wang}\ and\ \citenamefont
  {Truhlar}(2010)}]{wang2010including}%
  \BibitemOpen
  \bibfield  {author} {\bibinfo {author} {\bibfnamefont {B.}~\bibnamefont
  {Wang}}\ and\ \bibinfo {author} {\bibfnamefont {D.~G.}\ \bibnamefont
  {Truhlar}},\ }\href@noop {} {\bibfield  {journal} {\bibinfo  {journal} {J.
  Chem. Theory Comput.}\ }\textbf {\bibinfo {volume} {6}},\ \bibinfo {pages}
  {3330} (\bibinfo {year} {2010})}\BibitemShut {NoStop}%
\bibitem [{\citenamefont {Piquemal}, \citenamefont {Gresh},\ and\ \citenamefont
  {Giessner-Prettre}(2003)}]{piquemal2003improved}%
  \BibitemOpen
  \bibfield  {author} {\bibinfo {author} {\bibfnamefont {J.-P.}\ \bibnamefont
  {Piquemal}}, \bibinfo {author} {\bibfnamefont {N.}~\bibnamefont {Gresh}}, \
  and\ \bibinfo {author} {\bibfnamefont {C.}~\bibnamefont {Giessner-Prettre}},\
  }\href@noop {} {\bibfield  {journal} {\bibinfo  {journal} {J.~Phys.~Chem.~A}\
  }\textbf {\bibinfo {volume} {107}},\ \bibinfo {pages} {10353} (\bibinfo
  {year} {2003})}\BibitemShut {NoStop}%
\bibitem [{\citenamefont {Wang}\ \emph {et~al.}(2015)\citenamefont {Wang},
  \citenamefont {Rackers}, \citenamefont {He}, \citenamefont {Qi},
  \citenamefont {Narth}, \citenamefont {Lagardere}, \citenamefont {Gresh},
  \citenamefont {Ponder}, \citenamefont {Piquemal},\ and\ \citenamefont
  {Ren}}]{wang2015general}%
  \BibitemOpen
  \bibfield  {author} {\bibinfo {author} {\bibfnamefont {Q.}~\bibnamefont
  {Wang}}, \bibinfo {author} {\bibfnamefont {J.~A.}\ \bibnamefont {Rackers}},
  \bibinfo {author} {\bibfnamefont {C.}~\bibnamefont {He}}, \bibinfo {author}
  {\bibfnamefont {R.}~\bibnamefont {Qi}}, \bibinfo {author} {\bibfnamefont
  {C.}~\bibnamefont {Narth}}, \bibinfo {author} {\bibfnamefont
  {L.}~\bibnamefont {Lagardere}}, \bibinfo {author} {\bibfnamefont
  {N.}~\bibnamefont {Gresh}}, \bibinfo {author} {\bibfnamefont {J.~W.}\
  \bibnamefont {Ponder}}, \bibinfo {author} {\bibfnamefont {J.-P.}\
  \bibnamefont {Piquemal}}, \ and\ \bibinfo {author} {\bibfnamefont
  {P.}~\bibnamefont {Ren}},\ }\href@noop {} {\bibfield  {journal} {\bibinfo
  {journal} {J. Chem. Theory Comput.}\ }\textbf {\bibinfo {volume} {11}},\
  \bibinfo {pages} {2609} (\bibinfo {year} {2015})}\BibitemShut {NoStop}%
\bibitem [{\citenamefont {Narth}\ \emph {et~al.}(2016)\citenamefont {Narth},
  \citenamefont {Lagard{\`e}re}, \citenamefont {Polack}, \citenamefont {Gresh},
  \citenamefont {Wang}, \citenamefont {Bell}, \citenamefont {Rackers},
  \citenamefont {Ponder}, \citenamefont {Ren},\ and\ \citenamefont
  {Piquemal}}]{narth2016scalable}%
  \BibitemOpen
  \bibfield  {author} {\bibinfo {author} {\bibfnamefont {C.}~\bibnamefont
  {Narth}}, \bibinfo {author} {\bibfnamefont {L.}~\bibnamefont
  {Lagard{\`e}re}}, \bibinfo {author} {\bibfnamefont {E.}~\bibnamefont
  {Polack}}, \bibinfo {author} {\bibfnamefont {N.}~\bibnamefont {Gresh}},
  \bibinfo {author} {\bibfnamefont {Q.}~\bibnamefont {Wang}}, \bibinfo {author}
  {\bibfnamefont {D.~R.}\ \bibnamefont {Bell}}, \bibinfo {author}
  {\bibfnamefont {J.~A.}\ \bibnamefont {Rackers}}, \bibinfo {author}
  {\bibfnamefont {J.~W.}\ \bibnamefont {Ponder}}, \bibinfo {author}
  {\bibfnamefont {P.~Y.}\ \bibnamefont {Ren}}, \ and\ \bibinfo {author}
  {\bibfnamefont {J.-P.}\ \bibnamefont {Piquemal}},\ }\href@noop {} {\bibfield
  {journal} {\bibinfo  {journal} {J. Comput. Chem.}\ }\textbf {\bibinfo
  {volume} {37}},\ \bibinfo {pages} {494} (\bibinfo {year} {2016})}\BibitemShut
  {NoStop}%
\bibitem [{\citenamefont {Rackers}\ \emph {et~al.}(2017)\citenamefont
  {Rackers}, \citenamefont {Wang}, \citenamefont {Liu}, \citenamefont
  {Piquemal}, \citenamefont {Ren},\ and\ \citenamefont
  {Ponder}}]{rackers2017optimized}%
  \BibitemOpen
  \bibfield  {author} {\bibinfo {author} {\bibfnamefont {J.~A.}\ \bibnamefont
  {Rackers}}, \bibinfo {author} {\bibfnamefont {Q.}~\bibnamefont {Wang}},
  \bibinfo {author} {\bibfnamefont {C.}~\bibnamefont {Liu}}, \bibinfo {author}
  {\bibfnamefont {J.-P.}\ \bibnamefont {Piquemal}}, \bibinfo {author}
  {\bibfnamefont {P.}~\bibnamefont {Ren}}, \ and\ \bibinfo {author}
  {\bibfnamefont {J.~W.}\ \bibnamefont {Ponder}},\ }\href@noop {} {\bibfield
  {journal} {\bibinfo  {journal} {Phys. Chem. Chem. Phys.}\ }\textbf {\bibinfo
  {volume} {19}},\ \bibinfo {pages} {276} (\bibinfo {year} {2017})}\BibitemShut
  {NoStop}%
\bibitem [{\citenamefont {Thole}(1981)}]{thole1981molecular}%
  \BibitemOpen
  \bibfield  {author} {\bibinfo {author} {\bibfnamefont {B.~T.}\ \bibnamefont
  {Thole}},\ }\href@noop {} {\bibfield  {journal} {\bibinfo  {journal} {Chem.
  Phys.}\ }\textbf {\bibinfo {volume} {59}},\ \bibinfo {pages} {341} (\bibinfo
  {year} {1981})}\BibitemShut {NoStop}%
\bibitem [{\citenamefont {Hermann}, \citenamefont {DiStasio~Jr},\ and\
  \citenamefont {Tkatchenko}(2017)}]{hermann2017first}%
  \BibitemOpen
  \bibfield  {author} {\bibinfo {author} {\bibfnamefont {J.}~\bibnamefont
  {Hermann}}, \bibinfo {author} {\bibfnamefont {R.~A.}\ \bibnamefont
  {DiStasio~Jr}}, \ and\ \bibinfo {author} {\bibfnamefont {A.}~\bibnamefont
  {Tkatchenko}},\ }\href@noop {} {\bibfield  {journal} {\bibinfo  {journal}
  {Chem. Rev}\ }\textbf {\bibinfo {volume} {117}},\ \bibinfo {pages} {4714}
  (\bibinfo {year} {2017})}\BibitemShut {NoStop}%
\bibitem [{\citenamefont {Tkatchenko}\ \emph {et~al.}(2012)\citenamefont
  {Tkatchenko}, \citenamefont {DiStasio~Jr}, \citenamefont {Car},\ and\
  \citenamefont {Scheffler}}]{tkatchenko2012accurate}%
  \BibitemOpen
  \bibfield  {author} {\bibinfo {author} {\bibfnamefont {A.}~\bibnamefont
  {Tkatchenko}}, \bibinfo {author} {\bibfnamefont {R.~A.}\ \bibnamefont
  {DiStasio~Jr}}, \bibinfo {author} {\bibfnamefont {R.}~\bibnamefont {Car}}, \
  and\ \bibinfo {author} {\bibfnamefont {M.}~\bibnamefont {Scheffler}},\
  }\href@noop {} {\bibfield  {journal} {\bibinfo  {journal} {Phys. Rev. Lett.}\
  }\textbf {\bibinfo {volume} {108}},\ \bibinfo {pages} {236402} (\bibinfo
  {year} {2012})}\BibitemShut {NoStop}%
\bibitem [{\citenamefont {Donchev}(2006)}]{donchev2006many}%
  \BibitemOpen
  \bibfield  {author} {\bibinfo {author} {\bibfnamefont {A.}~\bibnamefont
  {Donchev}},\ }\href@noop {} {\bibfield  {journal} {\bibinfo  {journal} {J.
  Chem. Phys.}\ }\textbf {\bibinfo {volume} {125}},\ \bibinfo {pages} {074713}
  (\bibinfo {year} {2006})}\BibitemShut {NoStop}%
\bibitem [{\citenamefont {van~der Walt}, \citenamefont {Colbert},\ and\
  \citenamefont {Varoquaux}(2011)}]{walt2011numpy}%
  \BibitemOpen
  \bibfield  {author} {\bibinfo {author} {\bibfnamefont {S.}~\bibnamefont
  {van~der Walt}}, \bibinfo {author} {\bibfnamefont {S.~C.}\ \bibnamefont
  {Colbert}}, \ and\ \bibinfo {author} {\bibfnamefont {G.}~\bibnamefont
  {Varoquaux}},\ }\href@noop {} {\bibfield  {journal} {\bibinfo  {journal}
  {Computing in Science \& Engineering}\ }\textbf {\bibinfo {volume} {13}},\
  \bibinfo {pages} {22} (\bibinfo {year} {2011})}\BibitemShut {NoStop}%
\bibitem [{\citenamefont {O'Boyle}\ \emph {et~al.}(2011)\citenamefont
  {O'Boyle}, \citenamefont {Banck}, \citenamefont {James}, \citenamefont
  {Morley}, \citenamefont {Vandermeersch},\ and\ \citenamefont
  {Hutchison}}]{o2011open}%
  \BibitemOpen
  \bibfield  {author} {\bibinfo {author} {\bibfnamefont {N.~M.}\ \bibnamefont
  {O'Boyle}}, \bibinfo {author} {\bibfnamefont {M.}~\bibnamefont {Banck}},
  \bibinfo {author} {\bibfnamefont {C.~A.}\ \bibnamefont {James}}, \bibinfo
  {author} {\bibfnamefont {C.}~\bibnamefont {Morley}}, \bibinfo {author}
  {\bibfnamefont {T.}~\bibnamefont {Vandermeersch}}, \ and\ \bibinfo {author}
  {\bibfnamefont {G.~R.}\ \bibnamefont {Hutchison}},\ }\href@noop {} {\bibfield
   {journal} {\bibinfo  {journal} {J.~Cheminform.}\ }\textbf {\bibinfo {volume}
  {3}},\ \bibinfo {pages} {33} (\bibinfo {year} {2011})}\BibitemShut {NoStop}%
\bibitem [{\citenamefont {Faber}\ \emph {et~al.}(2017)\citenamefont {Faber},
  \citenamefont {Hutchison}, \citenamefont {Huang}, \citenamefont {Gilmer},
  \citenamefont {Schoenholz}, \citenamefont {Dahl}, \citenamefont {Vinyals},
  \citenamefont {Kearnes}, \citenamefont {Riley},\ and\ \citenamefont {von
  Lilienfeld}}]{faber2017prediction}%
  \BibitemOpen
  \bibfield  {author} {\bibinfo {author} {\bibfnamefont {F.~A.}\ \bibnamefont
  {Faber}}, \bibinfo {author} {\bibfnamefont {L.}~\bibnamefont {Hutchison}},
  \bibinfo {author} {\bibfnamefont {B.}~\bibnamefont {Huang}}, \bibinfo
  {author} {\bibfnamefont {J.}~\bibnamefont {Gilmer}}, \bibinfo {author}
  {\bibfnamefont {S.~S.}\ \bibnamefont {Schoenholz}}, \bibinfo {author}
  {\bibfnamefont {G.~E.}\ \bibnamefont {Dahl}}, \bibinfo {author}
  {\bibfnamefont {O.}~\bibnamefont {Vinyals}}, \bibinfo {author} {\bibfnamefont
  {S.}~\bibnamefont {Kearnes}}, \bibinfo {author} {\bibfnamefont {P.~F.}\
  \bibnamefont {Riley}}, \ and\ \bibinfo {author} {\bibfnamefont {O.~A.}\
  \bibnamefont {von Lilienfeld}},\ }\href@noop {} {\bibfield  {journal}
  {\bibinfo  {journal} {J. Chem. Theory Comput.}\ }\textbf {\bibinfo {volume}
  {13}},\ \bibinfo {pages} {5255} (\bibinfo {year} {2017})}\BibitemShut
  {NoStop}%
\bibitem [{\citenamefont {Glielmo}, \citenamefont {Sollich},\ and\
  \citenamefont {De~Vita}(2017)}]{glielmo2017accurate}%
  \BibitemOpen
  \bibfield  {author} {\bibinfo {author} {\bibfnamefont {A.}~\bibnamefont
  {Glielmo}}, \bibinfo {author} {\bibfnamefont {P.}~\bibnamefont {Sollich}}, \
  and\ \bibinfo {author} {\bibfnamefont {A.}~\bibnamefont {De~Vita}},\
  }\href@noop {} {\bibfield  {journal} {\bibinfo  {journal} {Phys.~Rev.~B}\
  }\textbf {\bibinfo {volume} {95}},\ \bibinfo {pages} {214302} (\bibinfo
  {year} {2017})}\BibitemShut {NoStop}%
\bibitem [{\citenamefont {Grisafi}\ \emph {et~al.}(2017)\citenamefont
  {Grisafi}, \citenamefont {Wilkins}, \citenamefont {Cs{\'a}nyi},\ and\
  \citenamefont {Ceriotti}}]{grisafi2017symmetry}%
  \BibitemOpen
  \bibfield  {author} {\bibinfo {author} {\bibfnamefont {A.}~\bibnamefont
  {Grisafi}}, \bibinfo {author} {\bibfnamefont {D.~M.}\ \bibnamefont
  {Wilkins}}, \bibinfo {author} {\bibfnamefont {G.}~\bibnamefont {Cs{\'a}nyi}},
  \ and\ \bibinfo {author} {\bibfnamefont {M.}~\bibnamefont {Ceriotti}},\
  }\href@noop {} {\bibfield  {journal} {\bibinfo  {journal} {arXiv preprint
  arXiv:1709.06757}\ } (\bibinfo {year} {2017})}\BibitemShut {NoStop}%
\bibitem [{\citenamefont {DiStasio~Jr}, \citenamefont {Gobre},\ and\
  \citenamefont {Tkatchenko}(2014)}]{distasio2014many}%
  \BibitemOpen
  \bibfield  {author} {\bibinfo {author} {\bibfnamefont {R.~A.}\ \bibnamefont
  {DiStasio~Jr}}, \bibinfo {author} {\bibfnamefont {V.~V.}\ \bibnamefont
  {Gobre}}, \ and\ \bibinfo {author} {\bibfnamefont {A.}~\bibnamefont
  {Tkatchenko}},\ }\href@noop {} {\bibfield  {journal} {\bibinfo  {journal} {J.
  Phys.: Condens. Matter}\ }\textbf {\bibinfo {volume} {26}},\ \bibinfo {pages}
  {213202} (\bibinfo {year} {2014})}\BibitemShut {NoStop}%
\bibitem [{\citenamefont {Wales}\ and\ \citenamefont
  {Doye}(1997)}]{wales1997global}%
  \BibitemOpen
  \bibfield  {author} {\bibinfo {author} {\bibfnamefont {D.~J.}\ \bibnamefont
  {Wales}}\ and\ \bibinfo {author} {\bibfnamefont {J.~P.}\ \bibnamefont
  {Doye}},\ }\href@noop {} {\bibfield  {journal} {\bibinfo  {journal}
  {J.~Phys.~Chem.~A}\ }\textbf {\bibinfo {volume} {101}},\ \bibinfo {pages}
  {5111} (\bibinfo {year} {1997})}\BibitemShut {NoStop}%
\bibitem [{\citenamefont {Wales}(2003)}]{wales2003energy}%
  \BibitemOpen
  \bibfield  {author} {\bibinfo {author} {\bibfnamefont {D.~J.}\ \bibnamefont
  {Wales}},\ }\href@noop {} {\emph {\bibinfo {title} {Energy landscapes:
  Applications to clusters, biomolecules and glasses}}}\ (\bibinfo  {publisher}
  {Cambridge University Press},\ \bibinfo {year} {2003})\BibitemShut {NoStop}%
\bibitem [{\citenamefont {Gr{\'a}fov{\'a}}\ \emph {et~al.}(2010)\citenamefont
  {Gr{\'a}fov{\'a}}, \citenamefont {Pitonak}, \citenamefont {Rezac},\ and\
  \citenamefont {Hobza}}]{grafova2010comparative}%
  \BibitemOpen
  \bibfield  {author} {\bibinfo {author} {\bibfnamefont {L.}~\bibnamefont
  {Gr{\'a}fov{\'a}}}, \bibinfo {author} {\bibfnamefont {M.}~\bibnamefont
  {Pitonak}}, \bibinfo {author} {\bibfnamefont {J.}~\bibnamefont {Rezac}}, \
  and\ \bibinfo {author} {\bibfnamefont {P.}~\bibnamefont {Hobza}},\
  }\href@noop {} {\bibfield  {journal} {\bibinfo  {journal} {J. Chem. Theory
  Comput.}\ }\textbf {\bibinfo {volume} {6}},\ \bibinfo {pages} {2365}
  (\bibinfo {year} {2010})}\BibitemShut {NoStop}%
\bibitem [{\citenamefont {Ambrosetti}\ \emph
  {et~al.}(2014{\natexlab{a}})\citenamefont {Ambrosetti}, \citenamefont
  {Alf\`e}, \citenamefont {DiStasio~Jr},\ and\ \citenamefont
  {Tkatchenko}}]{ambrosetti2014hard}%
  \BibitemOpen
  \bibfield  {author} {\bibinfo {author} {\bibfnamefont {A.}~\bibnamefont
  {Ambrosetti}}, \bibinfo {author} {\bibfnamefont {D.}~\bibnamefont {Alf\`e}},
  \bibinfo {author} {\bibfnamefont {R.~A.}\ \bibnamefont {DiStasio~Jr}}, \ and\
  \bibinfo {author} {\bibfnamefont {A.}~\bibnamefont {Tkatchenko}},\
  }\href@noop {} {\bibfield  {journal} {\bibinfo  {journal}
  {J.~Phys.~Chem.~Lett.}\ }\textbf {\bibinfo {volume} {5}},\ \bibinfo {pages}
  {849} (\bibinfo {year} {2014}{\natexlab{a}})}\BibitemShut {NoStop}%
\bibitem [{\citenamefont {Rez{\'a}c}, \citenamefont {Riley},\ and\
  \citenamefont {Hobza}(2011)}]{rezac2011extensions}%
  \BibitemOpen
  \bibfield  {author} {\bibinfo {author} {\bibfnamefont {J.}~\bibnamefont
  {Rez{\'a}c}}, \bibinfo {author} {\bibfnamefont {K.~E.}\ \bibnamefont
  {Riley}}, \ and\ \bibinfo {author} {\bibfnamefont {P.}~\bibnamefont
  {Hobza}},\ }\href@noop {} {\bibfield  {journal} {\bibinfo  {journal} {J.
  Chem. Theory Comput.}\ }\textbf {\bibinfo {volume} {7}},\ \bibinfo {pages}
  {3466} (\bibinfo {year} {2011})}\BibitemShut {NoStop}%
\bibitem [{\citenamefont {Burns}\ \emph {et~al.}(2017)\citenamefont {Burns},
  \citenamefont {Faver}, \citenamefont {Zheng}, \citenamefont {Marshall},
  \citenamefont {Smith}, \citenamefont {Vanommeslaeghe}, \citenamefont
  {MacKerell~Jr.}, \citenamefont {Merz~Jr.},\ and\ \citenamefont
  {Sherrill}}]{sherrill2017biofragment}%
  \BibitemOpen
  \bibfield  {author} {\bibinfo {author} {\bibfnamefont {L.~A.}\ \bibnamefont
  {Burns}}, \bibinfo {author} {\bibfnamefont {J.~C.}\ \bibnamefont {Faver}},
  \bibinfo {author} {\bibfnamefont {Z.}~\bibnamefont {Zheng}}, \bibinfo
  {author} {\bibfnamefont {M.~S.}\ \bibnamefont {Marshall}}, \bibinfo {author}
  {\bibfnamefont {D.~G.~A.}\ \bibnamefont {Smith}}, \bibinfo {author}
  {\bibfnamefont {K.}~\bibnamefont {Vanommeslaeghe}}, \bibinfo {author}
  {\bibfnamefont {A.~D.}\ \bibnamefont {MacKerell~Jr.}}, \bibinfo {author}
  {\bibfnamefont {K.~M.}\ \bibnamefont {Merz~Jr.}}, \ and\ \bibinfo {author}
  {\bibfnamefont {D.}~\bibnamefont {Sherrill}},\ }\href@noop {} {\bibfield
  {journal} {\bibinfo  {journal} {J. Chem. Phys.}\ }\textbf {\bibinfo {volume}
  {147}},\ \bibinfo {pages} {161727} (\bibinfo {year} {2017})}\BibitemShut
  {NoStop}%
\bibitem [{\citenamefont {Temelso}, \citenamefont {Archer},\ and\ \citenamefont
  {Shields}(2011)}]{temelso2011benchmark}%
  \BibitemOpen
  \bibfield  {author} {\bibinfo {author} {\bibfnamefont {B.}~\bibnamefont
  {Temelso}}, \bibinfo {author} {\bibfnamefont {K.~A.}\ \bibnamefont {Archer}},
  \ and\ \bibinfo {author} {\bibfnamefont {G.~C.}\ \bibnamefont {Shields}},\
  }\href@noop {} {\bibfield  {journal} {\bibinfo  {journal} {J.~Phys.~Chem.~A}\
  }\textbf {\bibinfo {volume} {115}},\ \bibinfo {pages} {12034} (\bibinfo
  {year} {2011})}\BibitemShut {NoStop}%
\bibitem [{\citenamefont {Schweizer}\ and\ \citenamefont
  {Dunitz}(2006)}]{schweizer2006quantum}%
  \BibitemOpen
  \bibfield  {author} {\bibinfo {author} {\bibfnamefont {W.~B.}\ \bibnamefont
  {Schweizer}}\ and\ \bibinfo {author} {\bibfnamefont {J.~D.}\ \bibnamefont
  {Dunitz}},\ }\href@noop {} {\bibfield  {journal} {\bibinfo  {journal} {J.
  Chem. Theory Comput.}\ }\textbf {\bibinfo {volume} {2}},\ \bibinfo {pages}
  {288} (\bibinfo {year} {2006})}\BibitemShut {NoStop}%
\bibitem [{\citenamefont {Tapavicza}\ \emph {et~al.}(2007)\citenamefont
  {Tapavicza}, \citenamefont {Lin}, \citenamefont {von Lilienfeld},
  \citenamefont {Tavernelli}, \citenamefont {Coutinho-Neto},\ and\
  \citenamefont {Rothlisberger}}]{tapavicza2007weakly}%
  \BibitemOpen
  \bibfield  {author} {\bibinfo {author} {\bibfnamefont {E.}~\bibnamefont
  {Tapavicza}}, \bibinfo {author} {\bibfnamefont {I.-C.}\ \bibnamefont {Lin}},
  \bibinfo {author} {\bibfnamefont {O.~A.}\ \bibnamefont {von Lilienfeld}},
  \bibinfo {author} {\bibfnamefont {I.}~\bibnamefont {Tavernelli}}, \bibinfo
  {author} {\bibfnamefont {M.~D.}\ \bibnamefont {Coutinho-Neto}}, \ and\
  \bibinfo {author} {\bibfnamefont {U.}~\bibnamefont {Rothlisberger}},\
  }\href@noop {} {\bibfield  {journal} {\bibinfo  {journal} {J. Chem. Theory
  Comput.}\ }\textbf {\bibinfo {volume} {3}},\ \bibinfo {pages} {1673}
  (\bibinfo {year} {2007})}\BibitemShut {NoStop}%
\bibitem [{\citenamefont {von Lilienfeld}\ and\ \citenamefont
  {Tkatchenko}(2010)}]{von2010two}%
  \BibitemOpen
  \bibfield  {author} {\bibinfo {author} {\bibfnamefont {O.~A.}\ \bibnamefont
  {von Lilienfeld}}\ and\ \bibinfo {author} {\bibfnamefont {A.}~\bibnamefont
  {Tkatchenko}},\ }\href@noop {} {\bibfield  {journal} {\bibinfo  {journal}
  {J.~Chem.~Phys.}\ }\textbf {\bibinfo {volume} {132}},\ \bibinfo {pages}
  {234109} (\bibinfo {year} {2010})}\BibitemShut {NoStop}%
\bibitem [{\citenamefont {Ho}\ and\ \citenamefont
  {Rabitz}(1996)}]{ho1996general}%
  \BibitemOpen
  \bibfield  {author} {\bibinfo {author} {\bibfnamefont {T.-S.}\ \bibnamefont
  {Ho}}\ and\ \bibinfo {author} {\bibfnamefont {H.}~\bibnamefont {Rabitz}},\
  }\href@noop {} {\bibfield  {journal} {\bibinfo  {journal} {J. Chem. Phys.}\
  }\textbf {\bibinfo {volume} {104}},\ \bibinfo {pages} {2584} (\bibinfo {year}
  {1996})}\BibitemShut {NoStop}%
\bibitem [{\citenamefont {Unke}\ and\ \citenamefont
  {Meuwly}(2017)}]{unke2017toolkit}%
  \BibitemOpen
  \bibfield  {author} {\bibinfo {author} {\bibfnamefont {O.~T.}\ \bibnamefont
  {Unke}}\ and\ \bibinfo {author} {\bibfnamefont {M.}~\bibnamefont {Meuwly}},\
  }\href@noop {} {\bibfield  {journal} {\bibinfo  {journal} {J. Chem. Inf.
  Model.}\ }\textbf {\bibinfo {volume} {57}},\ \bibinfo {pages} {1923}
  (\bibinfo {year} {2017})}\BibitemShut {NoStop}%
\bibitem [{\citenamefont {Blood-Forsythe}\ \emph {et~al.}(2016)\citenamefont
  {Blood-Forsythe}, \citenamefont {Markovich}, \citenamefont {DiStasio~Jr},
  \citenamefont {Car},\ and\ \citenamefont
  {Aspuru-Guzik}}]{blood2016analytical}%
  \BibitemOpen
  \bibfield  {author} {\bibinfo {author} {\bibfnamefont {M.~A.}\ \bibnamefont
  {Blood-Forsythe}}, \bibinfo {author} {\bibfnamefont {T.}~\bibnamefont
  {Markovich}}, \bibinfo {author} {\bibfnamefont {R.~A.}\ \bibnamefont
  {DiStasio~Jr}}, \bibinfo {author} {\bibfnamefont {R.}~\bibnamefont {Car}}, \
  and\ \bibinfo {author} {\bibfnamefont {A.}~\bibnamefont {Aspuru-Guzik}},\
  }\href@noop {} {\bibfield  {journal} {\bibinfo  {journal} {Chem.~Sci.}\
  }\textbf {\bibinfo {volume} {7}},\ \bibinfo {pages} {1712} (\bibinfo {year}
  {2016})}\BibitemShut {NoStop}%
\bibitem [{\citenamefont {Ponder}\ \emph {et~al.}(2010)\citenamefont {Ponder},
  \citenamefont {Wu}, \citenamefont {Ren}, \citenamefont {Pande}, \citenamefont
  {Chodera}, \citenamefont {Schnieders}, \citenamefont {Haque}, \citenamefont
  {Mobley}, \citenamefont {Lambrecht}, \citenamefont {DiStasio~Jr} \emph
  {et~al.}}]{ponder2010current}%
  \BibitemOpen
  \bibfield  {author} {\bibinfo {author} {\bibfnamefont {J.~W.}\ \bibnamefont
  {Ponder}}, \bibinfo {author} {\bibfnamefont {C.}~\bibnamefont {Wu}}, \bibinfo
  {author} {\bibfnamefont {P.}~\bibnamefont {Ren}}, \bibinfo {author}
  {\bibfnamefont {V.~S.}\ \bibnamefont {Pande}}, \bibinfo {author}
  {\bibfnamefont {J.~D.}\ \bibnamefont {Chodera}}, \bibinfo {author}
  {\bibfnamefont {M.~J.}\ \bibnamefont {Schnieders}}, \bibinfo {author}
  {\bibfnamefont {I.}~\bibnamefont {Haque}}, \bibinfo {author} {\bibfnamefont
  {D.~L.}\ \bibnamefont {Mobley}}, \bibinfo {author} {\bibfnamefont {D.~S.}\
  \bibnamefont {Lambrecht}}, \bibinfo {author} {\bibfnamefont {R.~A.}\
  \bibnamefont {DiStasio~Jr}},  \emph {et~al.},\ }\href@noop {} {\bibfield
  {journal} {\bibinfo  {journal} {J. Phys. Chem. B}\ }\textbf {\bibinfo
  {volume} {114}},\ \bibinfo {pages} {2549} (\bibinfo {year}
  {2010})}\BibitemShut {NoStop}%
\bibitem [{\citenamefont {Rasmussen}\ and\ \citenamefont
  {Williams}(2006)}]{rasmussen2006gaussian}%
  \BibitemOpen
  \bibfield  {author} {\bibinfo {author} {\bibfnamefont {C.~E.}\ \bibnamefont
  {Rasmussen}}\ and\ \bibinfo {author} {\bibfnamefont {C.~K.}\ \bibnamefont
  {Williams}},\ }\href@noop {} {\emph {\bibinfo {title} {Gaussian processes for
  machine learning}}},\ Vol.~\bibinfo {volume} {1}\ (\bibinfo  {publisher} {MIT
  press Cambridge},\ \bibinfo {year} {2006})\BibitemShut {NoStop}%
\bibitem [{\citenamefont {Van~Vleet}, \citenamefont {Misquitta},\ and\
  \citenamefont {Schmidt}(2017)}]{van2017new}%
  \BibitemOpen
  \bibfield  {author} {\bibinfo {author} {\bibfnamefont {M.~J.}\ \bibnamefont
  {Van~Vleet}}, \bibinfo {author} {\bibfnamefont {A.~J.}\ \bibnamefont
  {Misquitta}}, \ and\ \bibinfo {author} {\bibfnamefont {J.~R.}\ \bibnamefont
  {Schmidt}},\ }\href {\doibase 10.1021/acs.jctc.7b00851} {\bibfield  {journal}
  {\bibinfo  {journal} {Journal of chemical theory and computation}\ }
  (\bibinfo {year} {2017}),\ 10.1021/acs.jctc.7b00851}\BibitemShut {NoStop}%
\bibitem [{\citenamefont {Chu}\ and\ \citenamefont
  {Dalgarno}(2004)}]{chu2004linear}%
  \BibitemOpen
  \bibfield  {author} {\bibinfo {author} {\bibfnamefont {X.}~\bibnamefont
  {Chu}}\ and\ \bibinfo {author} {\bibfnamefont {A.}~\bibnamefont {Dalgarno}},\
  }\href@noop {} {\bibfield  {journal} {\bibinfo  {journal} {J.~Chem.~Phys.}\
  }\textbf {\bibinfo {volume} {121}},\ \bibinfo {pages} {4083} (\bibinfo {year}
  {2004})}\BibitemShut {NoStop}%
\bibitem [{\citenamefont {Ambrosetti}\ \emph
  {et~al.}(2014{\natexlab{b}})\citenamefont {Ambrosetti}, \citenamefont
  {Reilly}, \citenamefont {DiStasio~Jr},\ and\ \citenamefont
  {Tkatchenko}}]{ambrosetti2014long}%
  \BibitemOpen
  \bibfield  {author} {\bibinfo {author} {\bibfnamefont {A.}~\bibnamefont
  {Ambrosetti}}, \bibinfo {author} {\bibfnamefont {A.~M.}\ \bibnamefont
  {Reilly}}, \bibinfo {author} {\bibfnamefont {R.~A.}\ \bibnamefont
  {DiStasio~Jr}}, \ and\ \bibinfo {author} {\bibfnamefont {A.}~\bibnamefont
  {Tkatchenko}},\ }\href@noop {} {\bibfield  {journal} {\bibinfo  {journal} {J.
  Chem. Phys.}\ }\textbf {\bibinfo {volume} {140}},\ \bibinfo {pages} {18A508}
  (\bibinfo {year} {2014}{\natexlab{b}})}\BibitemShut {NoStop}%
\end{thebibliography}%

\end{document}